\documentclass[12pt,preprint]{aastex}

\shorttitle{Collapsed Cores in Globular Clusters}
\shortauthors{Stumpf et al.}

\begin{document}


\title{The Search for Planetary Mass Companions to \\
    Field Brown Dwarfs with HST/NICMOS\altaffilmark{1}}


\author{M.B. Stumpf, W. Brandner, V. Joergens, Th. Henning}
\affil{Max-Planck-Institut f\"ur Astronomie, K\"onigstuhl 17, D-69117 Heidelberg,
    Germany}
\author{H. Bouy}
\affil{Centro de Astrobiolog\'\i a, INTA-CSIC, PO BOX 78, E-28691, Villanueva 
de la Ca\~nada, Madrid, Spain}

\author{R. K\"ohler}
\affil{ZAH, Landessternwarte, K\"onigstuhl, D-69117 Heidelberg, Germany}

\author{M. Kasper}
\affil{European Southern Observatory, Karl-Schwarzschild-Strasse 2, D-85748 Garching, Germany}

\email{stumpf@mpia-hd.mpg.de}


\altaffiltext{1}{This work is based on observations made with the NASA/ESA Hubble Space Telescope, obtained at the Space Telescope Science Institute (STScI) and associated with program GO-10208. STScI is operated by the Association of Universities for Research in Astronomy, Inc., under NASA contract NAS 5-26555.}


\begin{abstract}
We present the results of a high-resolution spectral differential imaging survey of 12 nearby, relatively young field L dwarfs ($\le$\,1 Gyr) carried out with HST/NICMOS to search for planetary mass companions at small physical separations from their host.
The survey resolved two brown dwarf binaries: the L dwarf system Kelu-1\,AB and the newly discovered L/T transition system 2MASS\,031059+164815\,AB. For both systems common proper motion has already been confirmed in follow-up observations which have been  published elsewhere. The derived separations of the binaries are smaller than 6 AU and consistent with previous brown dwarf binary statistics. Their mass ratios of \emph{q} $\ge$ 0.8 confirm the preference for equal mass systems similar to a large number of other surveys.

Furthermore, we found tentative evidence for a companion to the L4 dwarf 2MASSW\,033703-175807, straddling the brown dwarf/planetary mass boundary and revealing an uncommonly low mass ratio system (\emph{q} $\approx$ 0.2) compared to the vast majority of previously found brown dwarf binaries. With a derived minimum mass of 10 $M_\mathrm{Jup}$ to 15 $M_\mathrm{Jup}$ a planetary nature of the secondary cannot be ruled out yet. However, it seems more likely to be a very low mass brown dwarf secondary at the border of the spectral T/Y transition regime, primarily due to its similarities to recently found very cool T dwarfs. This would make it one of the closest resolved brown dwarf binaries (0.087\arcsec\,$\pm$\,0.015\arcsec, corresponding to 2.52\,$\pm$\,0.44\,AU at a distance of 29\,pc) with the coolest (T$_{\mathrm{eff}}$ $\approx$ 600\,-\,630 K) and least massive companion to any L or T dwarf. 
\end{abstract}





\section{Introduction}
\label{Motiv_SDI}
Since the discovery of the first extrasolar planet around 51 Pegasi by \citeauthor{Mayor95} in 1995, more than 480 extrasolar planets (exoplanets) in more than 410 planetary systems have been identified as of August 2010\,\footnote{http://exoplanet.eu/catalog.php}. However, most of these detections were made using indirect detection methods like radial velocity (RV) surveys, the transit method or microlensing events. Further, the RV technique can in general only provide minimum masses (M\,sin\,$i$) and thus limited knowledge of the astrophysical properties of the companion.
From the beginning, these exoplanets revealed a remarkable diversity in their physical and orbital characteristics like separation, eccentricities and masses and thus challenged the planet formation scenario. In particular, the period-mass distribution shows tendencies which are hard to explain within the framework of current models (e.\,g.\,massive "hot Jupiter" were found at very small orbits, even down to a semi-major axis of only 0.02\,AU). A revision of the standard scenario seemed to be required. 
However, the discovery space of RV searches does not yet cover planets at distances larger than about 5\,AU (for statistics see \citealt{Marcy03}). The question is, if the observed period-mass distribution is, at least partly, a selection effect (higher sensitivity at shorter orbital periods) and thus might provide only an incomplete picture of the overall planet population, or if giant planets are really rare at larger separations. In addition, the indirect RV and microlensing techniques do not enable any photometric or spectroscopic information of the exoplanet companions and their physical properties such as brightness, color, effective temperature or composition. 

Therefore, a lot of attention has been directed towards direct imaging of planetary mass objects (PMO) in the last years. A direct detection would enable extensive follow-up studies to characterize the planets, but the main challenge of this method is to deal with the huge brightness contrast and the small angular separation between the faint planet companion and its host star. With the possibility of high-resolution space-based observations and the improvement of adaptive optics (AO) systems on large ground-based telescopes, the necessary sub-arcsecond spatial resolution became achievable. In addition, planetary mass objects and brown dwarfs are hotter at younger ages and hence the brightness contrast between the primary and a possible companion will be much smaller. Thus, in order to push the direct detection threshold down to the planetary mass regime, systematic searches focused especially on young stars in nearby associations.

First extensive \emph{Hubble Space Telescope} (HST) programs to search for substellar companions to post-TTauri stars  \citep{Brand00}, as well as to very low mass (VLM) stars and brown dwarfs in the Pleiades open cluster \citep{Mart00} did not reveal any companion at intermediate orbits. Since then, several other extended HST surveys of free-floating ultra-cool dwarfs and brown dwarfs showed that roughly 20$\%$ of them have substellar companions. Almost all of them constitute close to equal-mass systems (e.\,g.\,\citealt{Bouy03, Reid01, Burgasser03}) and hence, rather resemble more massive, stellar binaries than planetary systems. Consequently, some interesting questions arose: \\[2mm]
\hspace*{2cm}\parbox{14cm}{\textsl{Do "single" brown dwarfs evolve similar to single stars?}}\\[1mm]
\hspace*{2cm}\parbox{14cm} {\textsl{Are they capable of forming planetary systems?}}\\[2mm]
The detection of circumstellar disks around young brown dwarfs revealed that they at least posses the raw material to form planets (e.\,g.\,\citealt{Apai02, Pascu03, Liu03, RayJay03}; later confirmed by results of e.\,g.\,\citealt{Apai05, Luhman06, Luhman08_02, Scholz08}, see also reviews in \citealt{Luhman07} and \citealt{Henning08}).\\
On that account, we started in 2004 the first spectral differential imaging (SDI) survey with HST to directly detect planetary mass objects in close orbits around nearby free-floating brown dwarfs.  A direct image detection would have important implications on our understanding of planetary mass objects, their origin, the formation of substellar bodies in general and binarity in the brown dwarf domain in particular.

Brown dwarf multiplicity properties exhibit significant differences in frequency, separation and mass ratio distribution compared to Sun-like stars. While the binary fraction for Sun-like stars is $\approx$\,57\% \citep{Duquennoy1991}, this fraction steadily decreases to $\approx$\,26\,-\,42\,\% for early- to mid-type M dwarfs (e.g. \citealp{Reid1997, Bergfors10}) down to $\approx$\,10\,-\,30\,\% for the very low mass stars and brown dwarfs (e.g. \citealp{Bouy03, Burgasser07_2, Joergens08, Goldman08}). The same steady decline is true for the semi-major axis distribution which is spread over a wide range of separations for G dwarfs with a peak around 30\,AU \citep{Duquennoy1991}, smaller separations for M dwarfs where the distribution peaks between 4 and 30 AU \citep{Fischer1992}, and a very narrow separation distribution for the brown dwarfs which peaks at 3\,-\,10\,AU (e.g.\,\citealt{Allen07}). The mass ratio distribution for very low mass stars and brown dwarfs reveals a clear tendency to equal mass system components with a distinct peak around \emph{q}\,$\sim$\,1. In contrast, low \emph{q} values are much more common at all separations for binary systems of solar-type stars (see review by \citealt{Burgasser07_3} and references therein).

Recent analysis of radial velocity surveys confirm a 6\,-\,8\,\% frequency of planetary mass companions to Sun-like stars (e.g. \citealt{Grether06} and \citealt{Mordasini09} with references therein) at relatively small separations which peak around 1 AU \citep{Udry07}. While commonly observed mass ratios between solar type stars and substellar (planetary) mass objects range between 1:1000\,-\,1:100 (cf.\,Extrasolar Planets Encyclopedia\,\footnote{http://exoplanet.eu/catalog.php})  one would expect a much higher mass ratio of 1:20\,-\,1:5 for a system consisting of a planetary mass object orbiting a brown dwarf due to the lower initial mass of the primary. \citet{Apai08} point out that depending on the planet formation process, the frequency for lower-mass planets is expected to be higher than for solar type stars. For the semi-major axis distribution of such a system one would also expect relatively small separations of a few AU due to the much smaller sizes of the circumstellar disks around young brown dwarfs compared to solar type stars.

\section{Observing strategy}
\label{Obs_strat_SDI}
Because of the higher achievable contrast and the lack of suitable AO reference stars, the observations were conducted from space with HST as explained in the following.
The observing strategy that suited best to reach our goal of detecting planetary mass objects around brown dwarfs and to overcome the remaining brightness contrast, was the spectral differential imaging (SDI) technique \citep{Smith87}. This technique takes advantage of the fact that a cool ($\le$\,1300\,K) giant planet or low-mass brown dwarf (i.\,e.\,T dwarf) is much fainter in certain molecular bands due to strong absorption than in the neighboring continuum. 
Already \citet{Allard95} stated that the spectral energy distribution (SED) of brown dwarfs is very peculiar. The molecular opacities, which globally define the continuum of brown dwarfs, cause the SED to peak around 1.1\,$\mu$m for solar metallicities, almost independent of their effective temperature. Further, the observed \emph{J}--\emph{K} colors for L dwarfs are red (about +\,2 mag), while they are blue (\,-\,0.1 to +\,0.4 mag) for the T dwarfs. This is caused by dust grain formation \citep{Allard01}. Therefore, the brightness difference of any L dwarf\, -- T dwarf pair is smallest around 1.1\,$\mu$m, making the \emph{J}\,-\,band the best spectral regime to detect T\,-\,type or even cooler companions to L dwarfs. 

\begin{figure}[ht]
  \centering
    \begin{minipage}[t]{8.1cm}
       \includegraphics[angle=90,width=\textwidth]{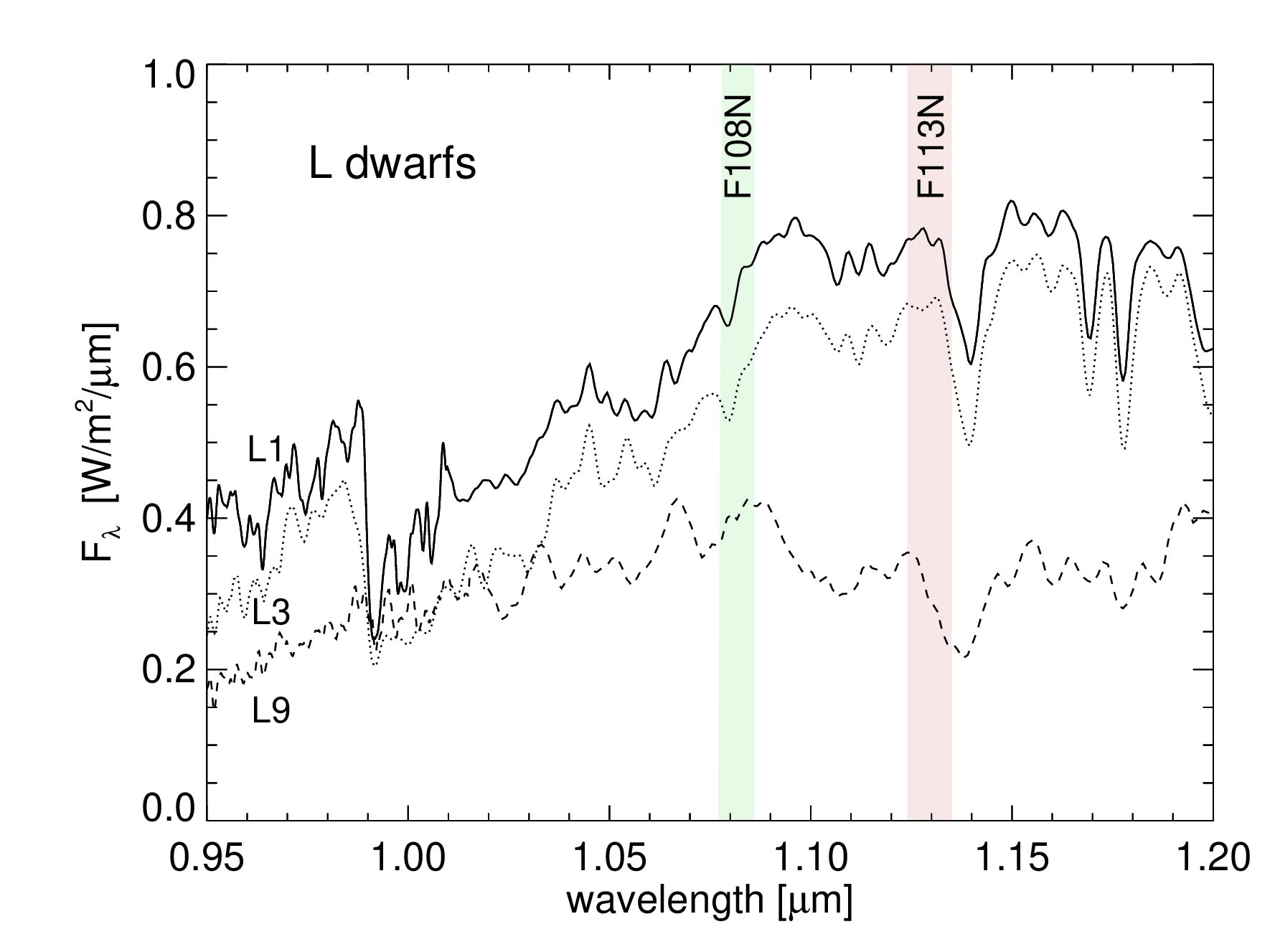}
     \end{minipage}
    \begin{minipage}[t]{8.1cm} 
       \includegraphics[angle=90,width=\textwidth]{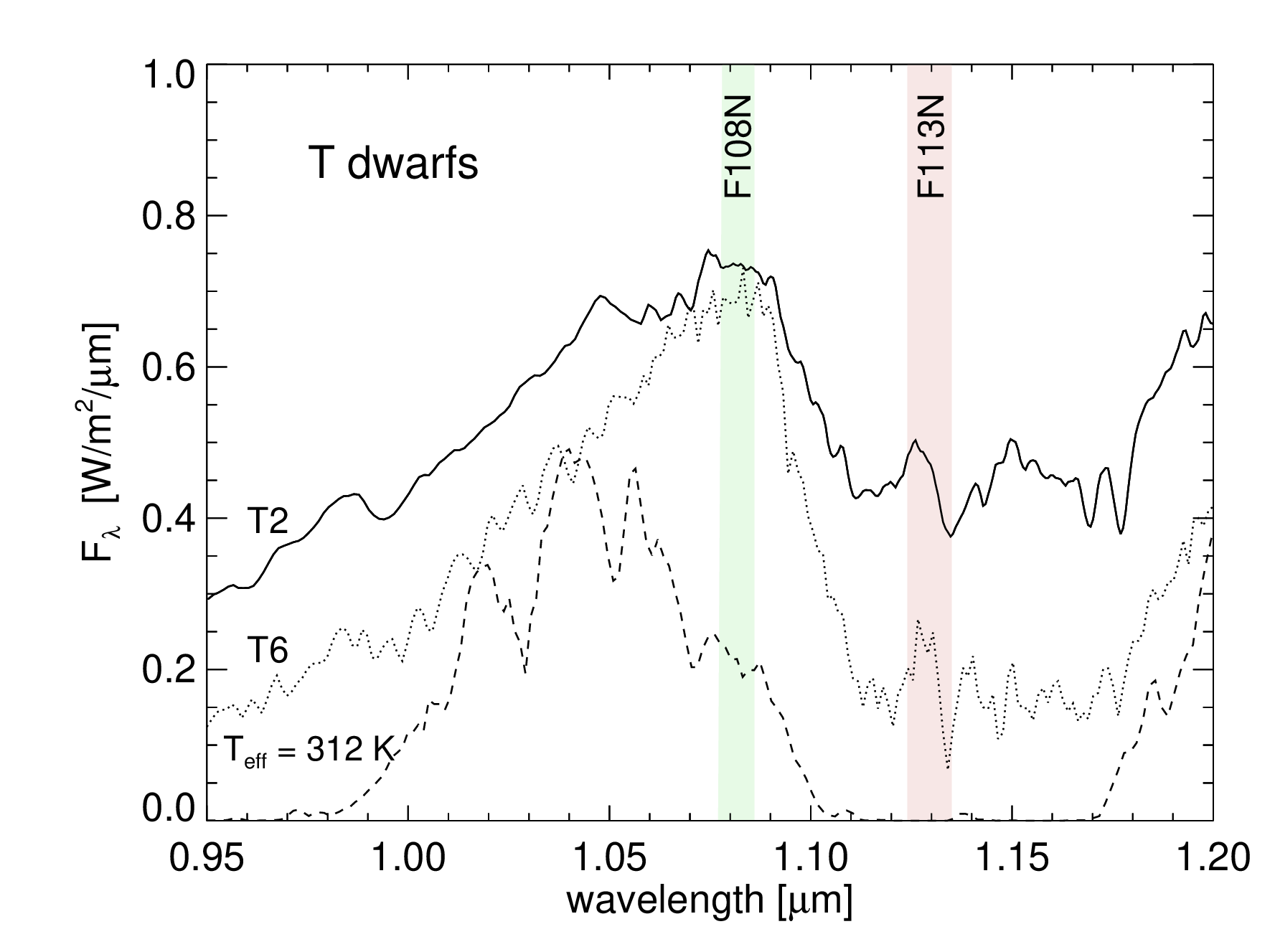}
      \end{minipage}
  \caption{ \footnotesize{Observed spectra of L and T dwarfs (from S. Leggett's online archive) with the superimposed bandpass of the HST F108N and F113N filters. The T dwarf plot additionally includes the synthetic spectrum of a 5\,$M_\mathrm{Jup}$, 1 Gyr old planetary mass object with T$_{\mathrm{eff}}$\,=\,312\,K \citep{Burrows03_2}. For a better comparison all flux is arbitrarily scaled.}}
   \label{SDI_bandpasses}
  \end{figure}

Figure~\ref{SDI_bandpasses} shows the spectral features of L and T dwarfs in this wavelength regime in detail. It reveals that L dwarfs have a comparative flat SED between 1.06\,$\mu$m and 1.15\,$\mu$m, while the flux of a T dwarf drops significantly at the beginning of the molecular water absorption band at $\sim$ 1.10\,$\mu$m. The T dwarf plot in this Figure~\ref{SDI_bandpasses} additionally includes a synthetic spectrum from \citet{Burrows03_2} for a 5\,$M_\mathrm{Jup}$, 1 Gyr old planetary mass object with T$_{\mathrm{eff}}$\,=\,312\,K and shows, that this effect is still present at these very low effective temperatures. Thus, a subtraction of two images obtained simultaneously, one in and one off, yet near this molecular band, will cancel out most of the equally bright PSF structure of the primary (L dwarf), revealing the much fainter signal from a cool, possibly planetary mass companion (see \citealt{Rosenthal96}; \citealt{Racine99}).
The narrow-band filters F108N and F113N of the NICMOS1 camera (NIC1) on board HST match particularly well the relevant bands (as shown in Figure~\ref{SDI_bandpasses}) and are ideally suited for the differential imaging technique.

\section{Sample selection}
Given the angular separation and luminosity ratio problem described above, the list of targets was assembled with the focus on brown dwarfs in proximity to the Sun and with a relatively young age. A closer distance to the Sun translates in a larger angular separation for a given physical separation between the brown dwarf and an eventual planetary-mass object, thus companions on smaller orbits can be detected. In addition, the detection threshold caused by sky background decreases for a fainter planetary mass object, since its apparent brightness becomes larger, the closer it is to the Sun.

Concerning the age, priority was given to targets with estimated ages younger than 1 Gyr. Giant planets, as well as young brown dwarfs are more luminous when they are young due to the remaining gravitational contraction energy from their formation process. Left without a resource of producing new energy they cool down rapidly and become dimmer with time. Based on evolutionary models and synthetic spectra, a young (100 Myr) giant planet or brown dwarf is $\sim$\,100\,-\,200 times more self-luminous than a 5 Gyr old one (e.\,g.\,\citealt{Burrows97}; \citealt{Baraffe03}). Hence any planetary mass companion to an old brown dwarf would be relatively cool and too faint for detection.

The age estimation of the targets in this survey was not straightforward since all of them are field brown dwarfs without any hints of being a member of an association or moving group. In general, \citet{Allen05} showed that proceeding down the spectral sequence from L to T dwarfs, the average age decreases and the relative proportion of young dwarfs ($\le$\,2 Gyr) increases if compared to M dwarfs. This is mainly caused by the lack of stable hydrogen burning in the brown dwarf core and therefore their rapid cooling through these temperature regimes with time \citep{Burrows01}. Our targets were selected according to spectral age indicators, such as the presence of lithium in the atmosphere. The detection and strength of lithium absorption is a clear confirmation of the substellar nature of these objects, as no lithium is expected for masses $\ge$ 65 $M_\mathrm{Jup}$ due to lithium destruction by proton capture in the deeper layers of the convection zone (e.g. \citealp{Bodenheimer1965, Nelson1993}). Combined with the still relatively high effective temperature representative for the spectral type L this is a clear sign of youth $\le$\,1 Gyr (\citealt{Basri98_2}; \citealt{Magazzu93}). Moreover, \citet{Gizis00} point out that chromospheric activity is primarily observed in older and hence stellar mass L dwarfs. They also find a clear anti-correlation between the presence of lithium and H$_\alpha$ emission. Thus, early L dwarfs with a very weak or non-existent H$_\alpha$ emission line are more likely young brown dwarfs rather than stars.

The final sample was compiled out of  $\sim$\,250 L dwarfs cataloged in D.\,Kirkpatrick's online archive\footnote{http://www.DwarfArchives.org} as of 2004 (start of this project) and consists of 12 young L dwarfs which fulfilled the following requirements:
\begin{itemize}
\item the object is within 30\,pc to the Sun
\item the \ion{Li}{1} absorption at 6708 \r{A} is present and therefore a clear confirmation of their substellar nature, and indication of youth.
\item lack of strong chromospheric activity (H$_\alpha$ emission) which is preferentially observed in stellar mass L dwarfs.
\item the objects are isolated in the sense of no known close companion so far.
\end{itemize}

These age and proximity criteria helped to enlarge the range of mass and separation over which the survey would be sensitive and to maximize the discovery probability. The properties of all selected targets are summarized in Table~\ref{Diff_imag_targets}. Although there are no Lithium measurements available for the two mid\,-\,L dwarfs 2MASSW\,004521+1634 and 2MASSI\,0835--0819, they were included since they belonged to the 10 closest L dwarfs known and because of their brightness compared to other L dwarfs of the same spectral type. A similar decision was made for Kelu-1 which had only a weak detection of the Lithium absorption line but was apparently overluminos \citep{Mart99_2, Leggett01, Goli04_2} and showed photometric and spectroscopic variability \citep{Clarke02, Clarke03}. Furthermore, the nearby T dwarf 2MASSW\,055919--1404 was added to calibrate the spectral differential imaging method for HST/NICMOS.

\section{Observations and SDI data reduction}
The observations of the twelve isolated L dwarfs and the calibration T dwarf were scheduled as HST program GO 10208 (PI: W. Brandner) from September 2004 to July 2005. The sources were observed with the NICMOS1 (NIC1) camera, providing a high-resolution pixel scale of 0\,\farcs0432 and a field of view (FoV) of 11\arcsec \,x 11\arcsec, in the two narrowband filters F108N (1.08 $\mu$m) and F113N (1.13 $\mu$m). The observation sequences\- consisted of 4 exposures at two different detector positions in each filter. This two\,-\,point dither pattern facilitates an optimized PSF sampling, effectively rejects bad pixels and gains redundancy against cosmic ray events. Altogether, two orbits per target were necessary to achieve the required S/N ratio of 6\,-\,8 for a limiting magnitude of \emph{J}\,=\,20 mag in the narrowband filters, and a brightness difference of $\approx$ 6 mag at a separation of 0.3$\arcsec$. All data were acquired in MULTIACCUM mode and the total integration times per filter were 2560\,s (F108N) and 2816\,s (F113N), respectively. In order to limit possible HST breathing variations in the PSF, differential imaging observations in the F108N and F113N filters were obtained in each of the orbits.

For the analysis of the data sets, the HST pipeline reduced images were used, followed by an additional bad pixel masking in order to compensate for the non-optimal bad pixel mask available in the pipeline. The SDI reduction was then accomplished with a custom made IDL program: First, a square aperture around the brown dwarf is extracted from all four images in each filter. To make use of the full integration time, the extracted frames were then co-added after they were aligned with respect to the first image of the related filter. This alignment was done with a 2D cross-correlation and a FFT-shift based combination, which provides a sub-pixel accuracy. In the next step, a sub-pixel re-sampling was performed in order to transform the observation in the F108N filter to the same $\lambda$/D scale as in the F113N filter (where $\lambda$ is the respective observing wavelength and \emph{D} is the HST primary mirror diameter). The final difference between the two filters was then calculated after additional alignment of the two master images with the shift algorithm, followed by a flux calibration to account for any flux losses during the previous re-scale process. In cases where a clear positive residual remained in the reduced image, the difference was also obtained separately for the individual dither positions (hence only two images per filter). This provides two independent detections of a real signal, or the exclusion, if it only exists at one position and is most likely caused by a cosmic ray remnant or hot pixel buried in the central core of the PSF. 

\section{Results}
\begin{figure}[ht]
    \setlength{\unitlength}{1cm}
  \begin{minipage}[t]{6.9cm}
        \includegraphics[width=\textwidth]{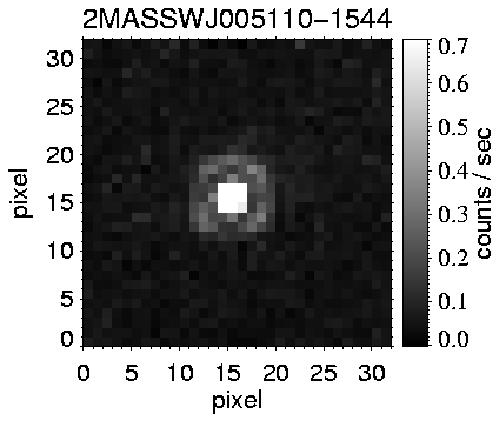}
    \end{minipage}\hfill
    \begin{minipage}[t]{6.9cm}
        \includegraphics[width=\textwidth]{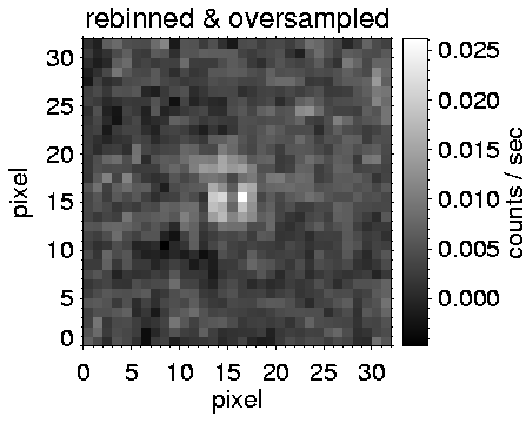}
     \end{minipage}\vspace{1.5ex}     
     \begin{minipage}[t]{6.9cm}
        \includegraphics[width=\textwidth]{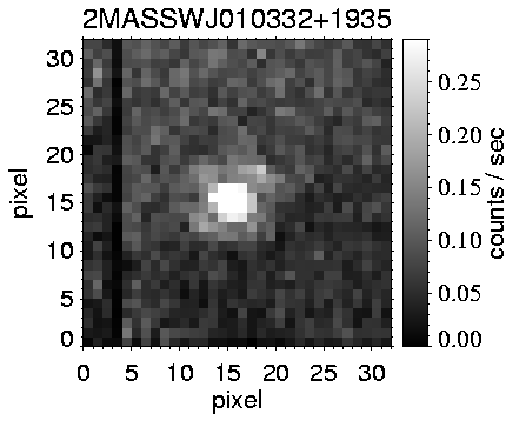}
    \end{minipage}\hfill
    \begin{minipage}[t]{6.9cm}
        \includegraphics[width=\textwidth]{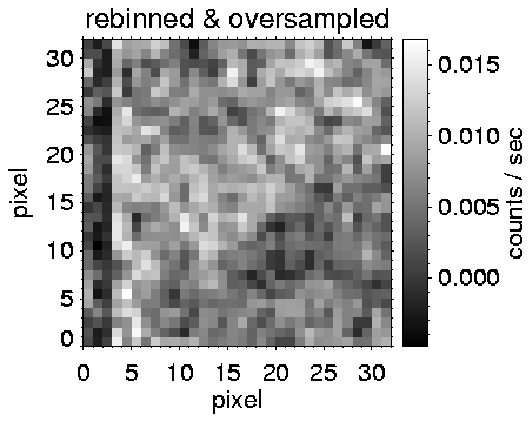}
     \end{minipage}\vspace{1.5ex}
     \begin{minipage}[t]{6.9cm}
        \includegraphics[width=\textwidth]{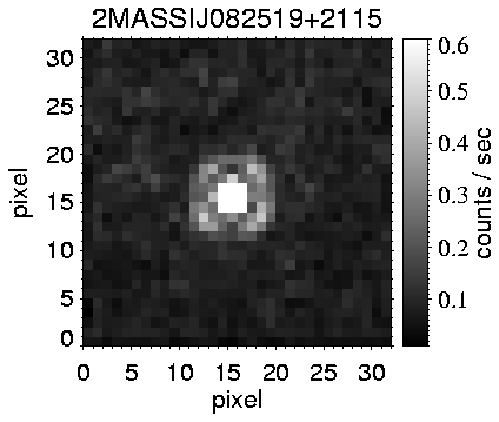}
    \end{minipage}\hfill
    \begin{minipage}[t]{6.9cm}
        \includegraphics[width=\textwidth]{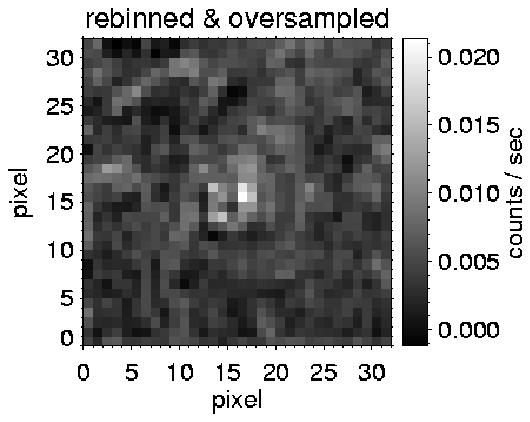}
     \end{minipage}  
    \caption{\footnotesize{Result images for five of the nine L dwarfs without any companion: The image on the left represents the added image of all 4 exposures taken in the F108N filter with the upper cut-level reduced for a better visibility of the background (thus not representing the real peak values). The right image displays the final reduced image after the subtraction of the two filters. The positive signals visible in the central part of the final images of 2MASS\,0051-1544 and 2MASS\,0825+2115, are not simultaneously present in each of the individual reduced detector position images and are therefore residuals from bad pixels buried in the original PSF. The overall smooth result images confirm the correctness of our reduction procedure.}}
   \label{Diff_imaging_result_1}
\end{figure}

\begin{figure}[ht]
    \setlength{\unitlength}{1cm}
  \begin{minipage}[t]{5.9cm}
        \includegraphics[width=\textwidth]{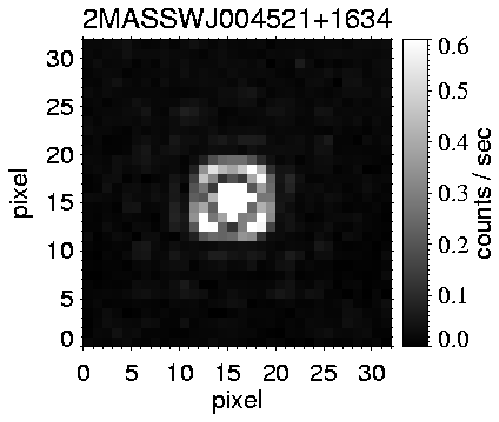}
    \end{minipage}\hfill
    \begin{minipage}[t]{5.9cm}
        \includegraphics[width=\textwidth]{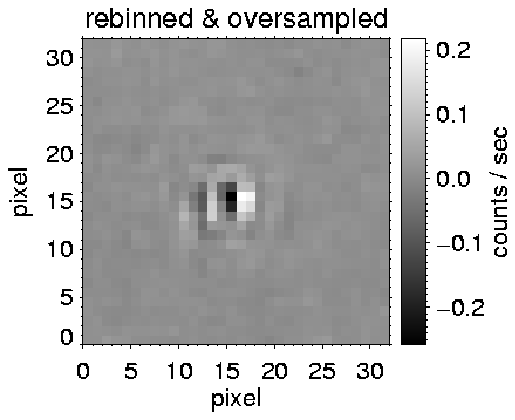}
     \end{minipage}\vspace{1.5ex}     
    \begin{minipage}[t]{5.9cm}
        \includegraphics[width=\textwidth]{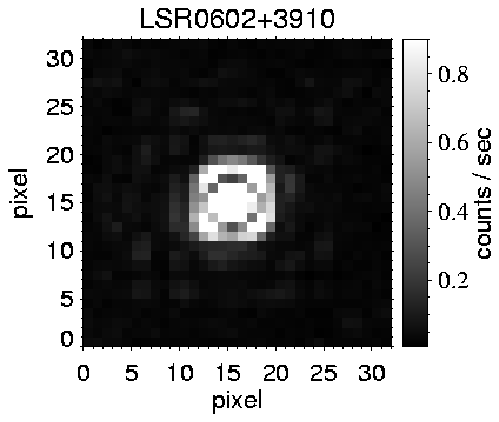}
    \end{minipage}\hfill
    \begin{minipage}[t]{5.9cm}
        \includegraphics[width=\textwidth]{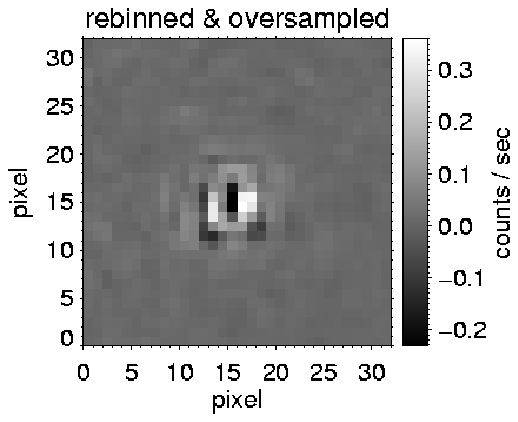}
     \end{minipage}\vspace{1.5ex}     
   \begin{minipage}[t]{5.9cm}
        \includegraphics[width=\textwidth]{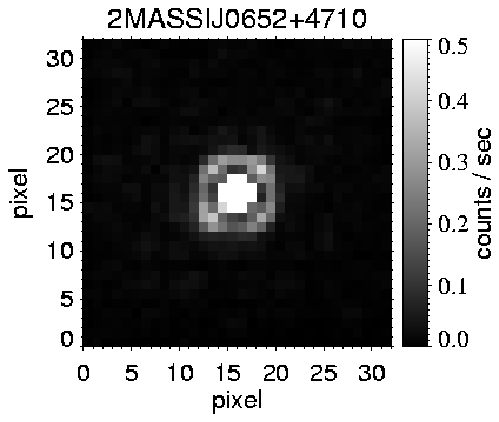}
    \end{minipage}\hfill
    \begin{minipage}[t]{5.9cm}
        \includegraphics[width=\textwidth]{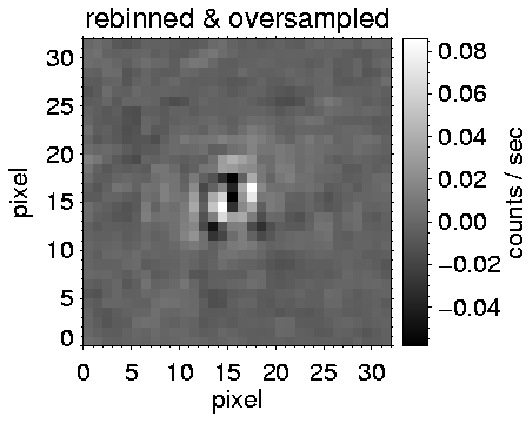}
     \end{minipage}\vspace{1.5ex}
     \begin{minipage}[t]{5.9cm}
        \includegraphics[width=\textwidth]{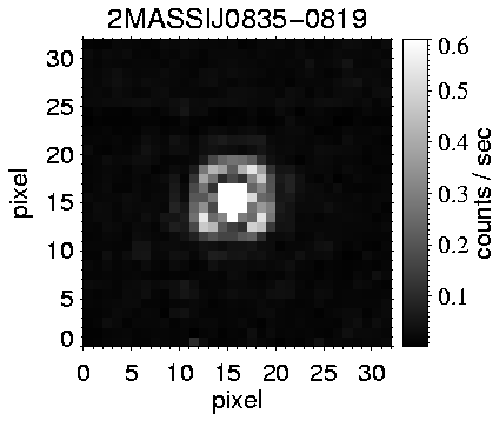}
    \end{minipage}\hfill
    \begin{minipage}[t]{5.9cm}
        \includegraphics[width=\textwidth]{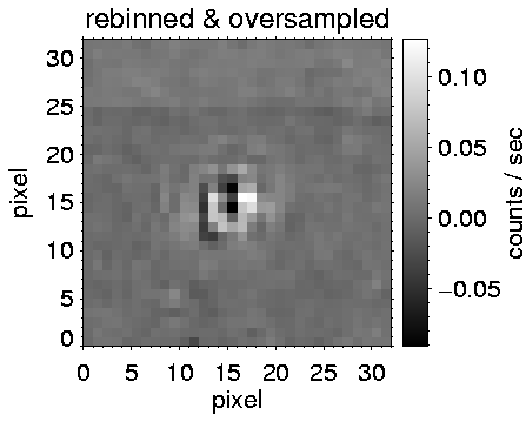}
     \end{minipage}\vspace{-1.8ex} 
 \caption{\footnotesize{same as Figure \protect \ref{Diff_imaging_result_1}. These are the four brightest and closest targets of the sample. The systematic mirrored dark and bright features are most probably caused by the systematic variations in the PSF structure due to differential aberrations of the filters. However, no other significant  positive signal can be found in the vicinity of the brown dwarfs.}}
   \label{Diff_imaging_result_2}
\end{figure}

The visual examination of the reduced images reveals that for nine of our targets the residuals are at the level expected from photon noise and no significant positive signal can be detected. 
 The final images of these data sets are shown in Figures \ref{Diff_imaging_result_1} and \ref{Diff_imaging_result_2}. The remaining single positive signals in some of the targets in Figure \ref{Diff_imaging_result_1}  just close to the original PSF center, are examples for residuals caused by bad pixels which could not be corrected by the additional bad pixel masking, since they were buried in the bright core of the PSF. They are not visible in both of the ``single" position SDI images simultaneously and thus do not correspond to a real signal of a possible faint companion. For the four brightest and closest sample targets 2MASSW\,0045+1634, LSR\,0602+3910, 2MASSI\,0652+4710 and 2MASSI\,0835-0819 (Figure \ref{Diff_imaging_result_2}) the final SDI results show systematic mirrored dark and bright features in the $\sim$ 6\,-\,7 central pixels of the images. These are common residuals in contrast-limited cases and are most probably caused by the systematic variations in the PSF structure due to differential aberrations of the filters. Since the other five targets are much fainter and their residuals are mostly dominated by photon noise, this pattern is no longer visible. If they were brighter their residuals would look the same. 
  
\begin{figure}[ht]
  \centering
       \begin{minipage}[t]{10cm}
       \includegraphics[angle=90,width=\textwidth]{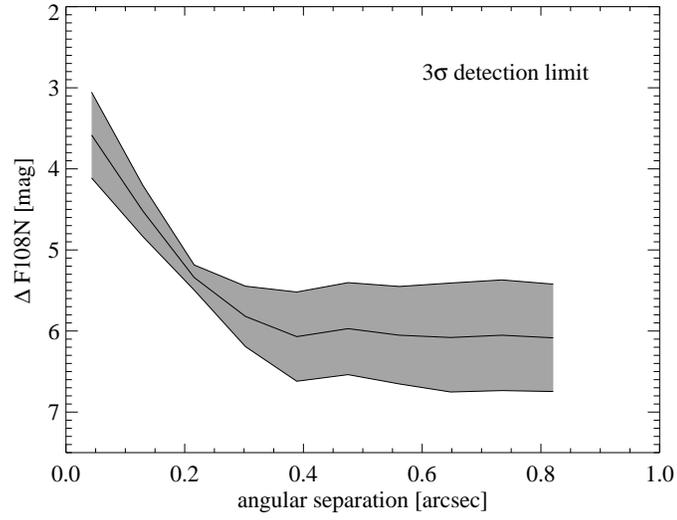}
     \end{minipage}

  \caption{ \footnotesize{The image illustrates the achievable sensitivity in magnitude difference $\Delta$\,F108N for our SDI survey as a function of angular separation. The bold solid line corresponds to the mean 3$\sigma$ detection limit of all observing runs (excluding the 2 confirmed and clearly resolved binary systems). The upper and lower shaded curves correspond to the standard deviation.}}
   \label{SDI_sensitivity_mag}
  \end{figure}

\begin{figure}[ht]
  \centering
    \begin{minipage}[t]{8.15cm}
       \includegraphics[angle=90,width=\textwidth]{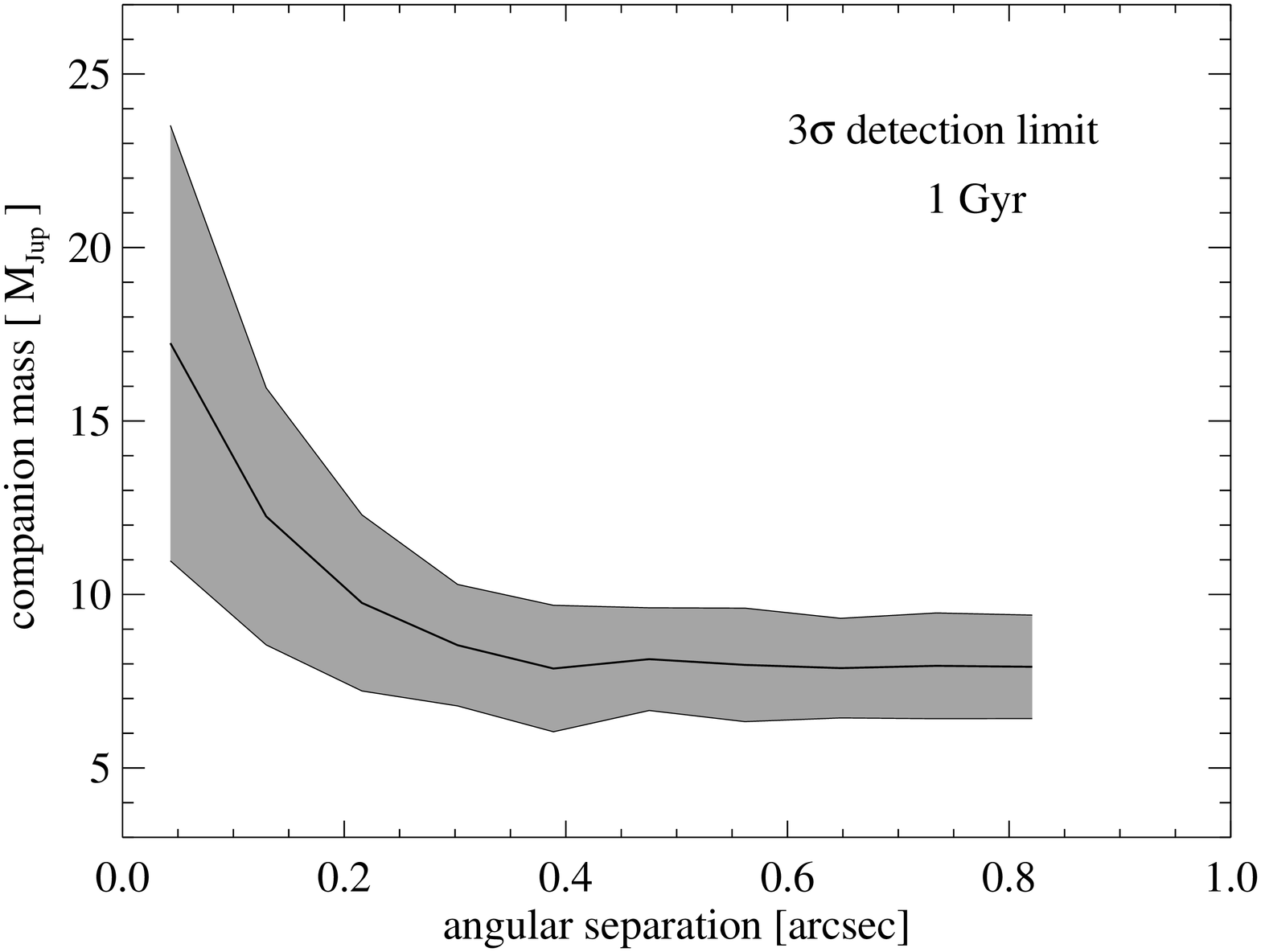}
     \end{minipage}
    \begin{minipage}[t]{8.15cm} 
       \includegraphics[angle=90,width=\textwidth]{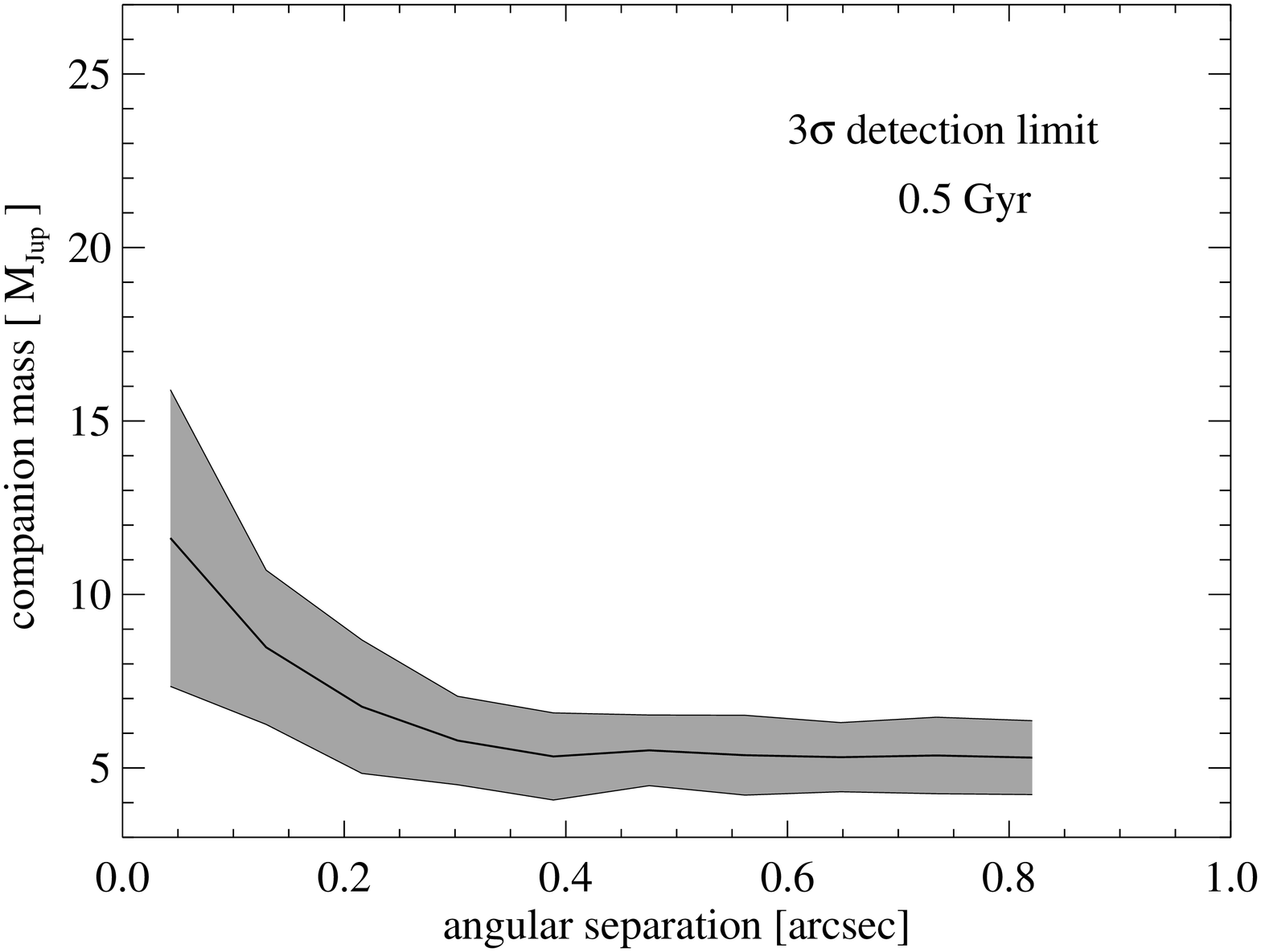}
      \end{minipage}
  \caption{ \footnotesize{Comparison of achieved detection limits for companion masses at ages of 1 Gyr (left image) and 0.5 Gyr (right image). The derived contrast limits of each observing run were converted into masses based on the evolutionary models from \citet{Baraffe03} as described in the text. The images display the mean detectable mass limit with the shaded areas representing the standard deviation.}}
   \label{SDI_sensitivity_mass}
  \end{figure}

To asses the overall sensitivity for the survey, the standard deviation of the background residuals after PSF subtraction was calculated for each data set as a function of radial separation from the host brown dwarf. The limits are then based on the median of the statistical errors at the various radii and the primary F108N magnitude. Figure~\ref{SDI_sensitivity_mag} shows the derived 3$\sigma$ narrowband detection limits. The bold solid line represents the mean of the data sets, while the shaded area corresponds to the standard deviation. At a separation of 0.05$\arcsec$ from the primary, a mean contrast of $\Delta$\,F108N\,=\,3.5 mag can already be achieved. For the brighter brown dwarfs in the survey we reach a contrast as large as $\sim$\,6.7 mag at separations $\ge$\,0.4$\arcsec$ from the primary. For the fainter brown dwarfs in the sample we still reach a contrast of $\sim$\,5.4 mag for separations $\ge$\,0.4$\arcsec$.
In order to determine the lower mass limit sensitivity of the survey, one can translate the contrast (in combination with the brightness of the primary) into a mass using the theoretical  COND evolutionary models from \citet{Baraffe03}. First of all, we converted the known \emph{J}-\,band magnitudes of the primary L dwarfs into \emph{Y}-\,band magnitudes. \citet{Hillenbrand02} and \citet{Hewett06} showed that the \emph{Y}\,--\,\emph{J} colors remain relatively constant at 1.0\,$\pm$\,0.15 mag from early-L through late-T type dwarfs. Thereafter, we applied the mass-luminosity relationships of the \emph{Y}-\,band (I.\,Baraffe, private communication) which are closest to the F108N bandpass. Figure~\ref{SDI_sensitivity_mass} illustrates the resulting mass limits for 1 Gyr (left image) and 0.5 Gyr (right image). Similar to Figure~\ref{SDI_sensitivity_mag}, the bold solid line corresponds to the mean value of the individual observing runs and the shaded areas represent the standard deviation.

Altogether, the smooth results after the SDI reduction indicate the absence of a planetary mass companion with a minimum mass of 6\,-\,11\,$M_\mathrm{Jup}$ at an age of 1 Gyr and between 0.07$\arcsec$ and $\sim$\,1.4$\arcsec$ around the brown dwarfs ($\sim$ 1 - 40 AU, depending on their distance to the Sun). This is an area usually most strongly affected by PSF residuals for non-differential imaging observations. For primaries with an age of 0.5 Gyr our observations even exclude any planetary mass object down to 5\,-\,7\,$M_\mathrm{Jup}$. This result validates our observing and data analysis strategy. For the three remaining L dwarfs, the observations revealed very interesting results which will be discussed in the following sections.

\subsection{Two newly resolved brown dwarf binaries}
\label{Diff_binaries}
\begin{figure}[ht]
  \centering
    \setlength{\unitlength}{1cm}
    \begin{minipage}[t]{5.5cm}
       \thicklines\framebox(5.5,5.5){\includegraphics[width=\textwidth]{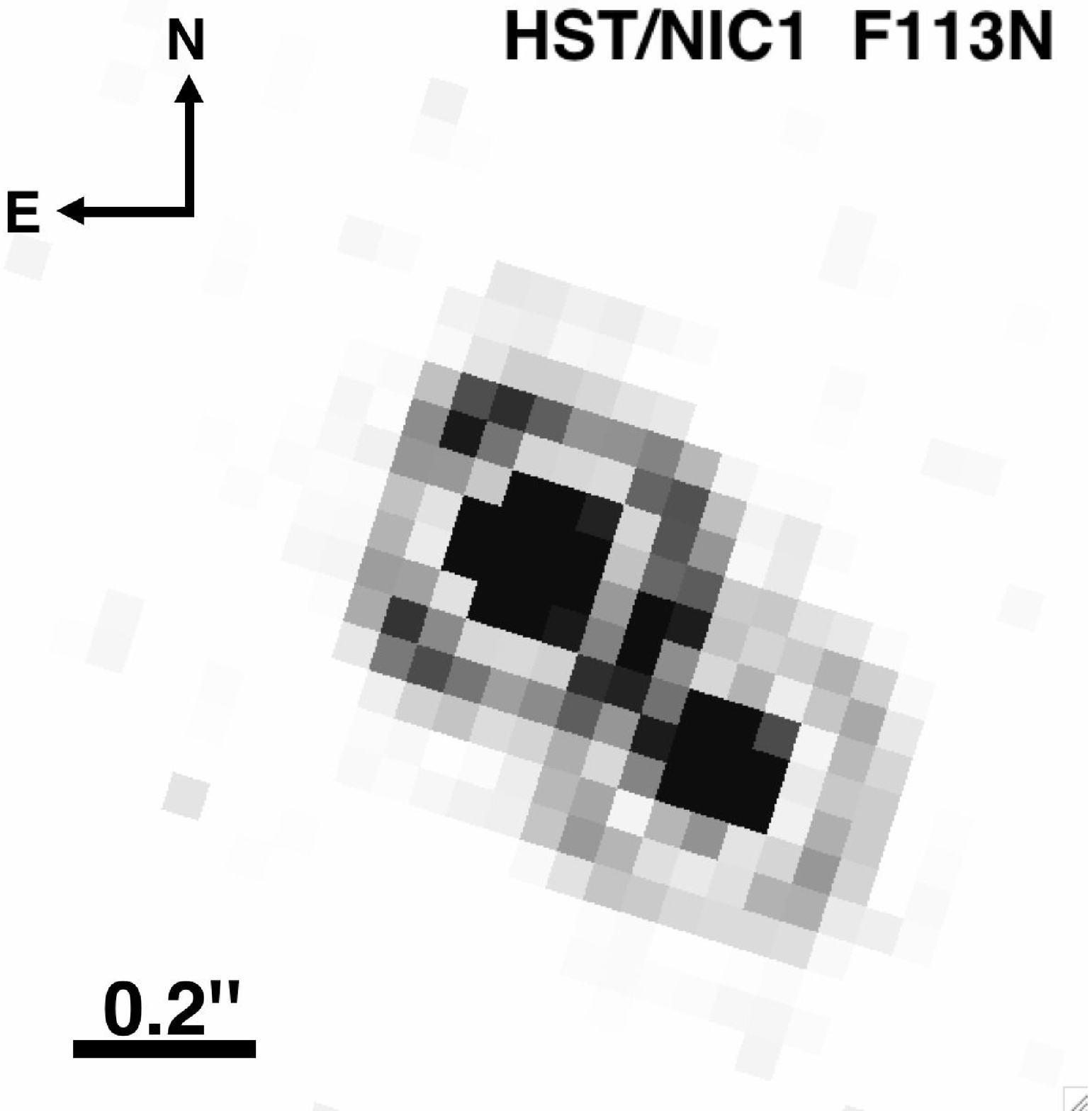}}
     \end{minipage}\hspace{1.5cm}
    \begin{minipage}[t]{5.5cm} 
      \thicklines\framebox(5.5,5.5){ \includegraphics[width=\textwidth]{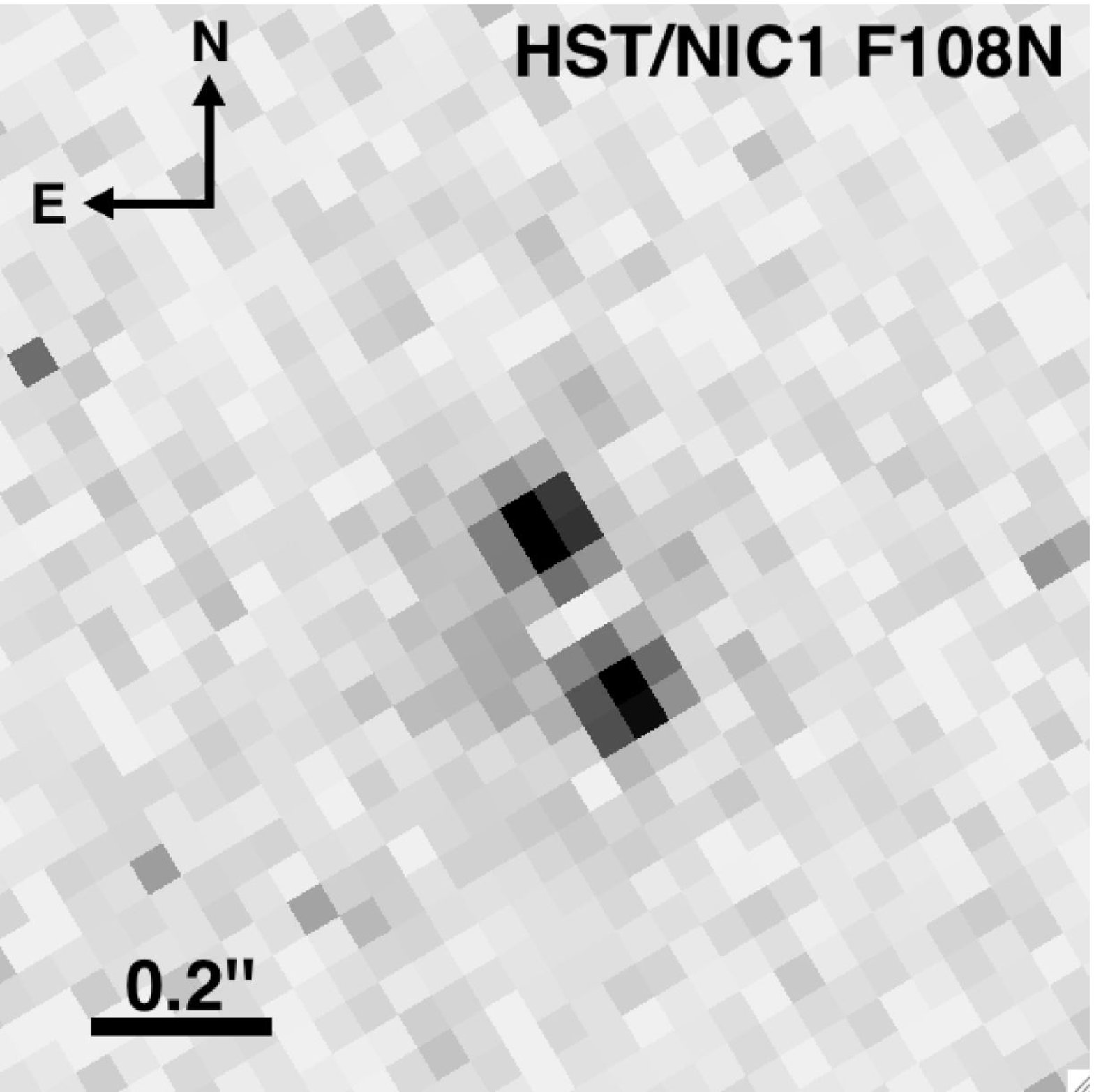}}
      \end{minipage}
  \caption{ \footnotesize{NICMOS images of the two resolved L dwarf binaries: on the left the bright, initially classified L2 dwarf Kelu-1\,AB, and on the right the much fainter L9 dwarf 2MASS\,0310+1648\,AB}}
   \label{HST_binaries}
  \end{figure}

Even though we tried to avoid known binaries during the sample selection process, the observations revealed that two of the targets are close brown dwarf binary systems. Figure \ref{HST_binaries} shows the NICMOS detection images where both systems are clearly resolved in two components. 

\subsubsection{Kelu-1\,AB}
Kelu-1 was discovered to be a binary system by \citet{Liu} and also independently resolved by \citet {Gelino}, both with AO from the ground and just four months before our own HST observation in July 2005. This solved the by then unexplained overluminosity of the ``single" object. The HST observation presented here, is now the first high-resolution confirmation of the physical association of both components, by proofing that they are a common proper motion pair. The system showed clear evidence for orbital motion with a significant increase in separation by 15\,mas in four months out to 299.8 $\pm$ 0.2\,mas and a position angle (PA) of 221.33$\degr$ $\pm$ 0.04$\degr$, hinting at  a short orbit binary. This makes Kelu-1\,AB a good target for a monitoring program. The extensive follow up observations derived with the \emph{Very Large Telescope} (VLT) and all results concerning this system are discussed in Stumpf et\,al.\,(\citeyear{Stumpf09}; 2010b, submitted).

\subsubsection{2MASSW\,0310+1648\,AB}
The second discovered very low mass binary is 2MASSW\,0310+1648\,AB. With an original spectral type of L9, this brown dwarf belongs to the still peculiar L/T transition objects which exhibit several unusual characteristics. Resolved for the first time  as a binary system with almost equally bright components, it adds another system to the already higher binary fraction among the brown dwarfs that span the transition between the L and T spectral class (e.g.\,\citealp{Burgasser06_1, Burgasser10_1}). The first astrometric measurements yield a separation of 204.3 $\pm$ 0.4\,mas and a PA of 206.4$\degr$ $\pm$ 0.1$\degr$ on 2004 September 24. Including follow up observations, this target is discussed in \citet{Stumpf10_1}.

\subsection{A planetary companion candidate}
\label{Diff_companion}
\begin{figure}[ht]
      \setlength{\unitlength}{1cm}
   \begin{picture}(14.5,18.5) 
   \put(-0.2,16.7){\begin{minipage}[t]{3cm}
     \includegraphics[width=\textwidth,angle=-63.5]{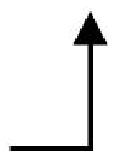}
     \end{minipage} }
   \put(4.5,12.2){\begin{minipage}[t]{6.9cm}
     \includegraphics[width=\textwidth]{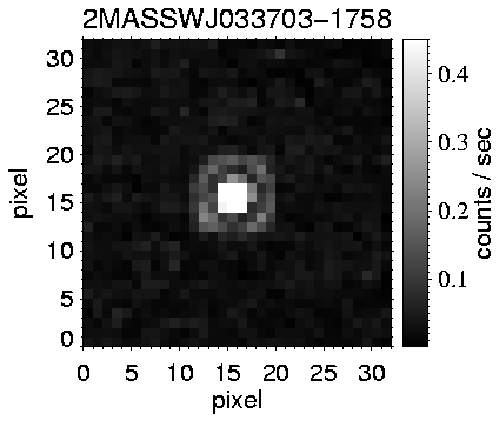}
     \end{minipage} }          
   \put(0.3,6.5){ \begin{minipage}[t]{6.9cm} 
     \includegraphics[width=\textwidth]{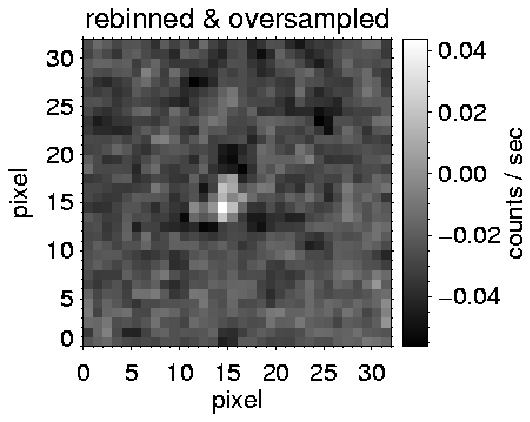}
     \end{minipage}}
    \put(7.6,6.5){ \begin{minipage}[t]{6.9cm} 
        \includegraphics[width=\textwidth]{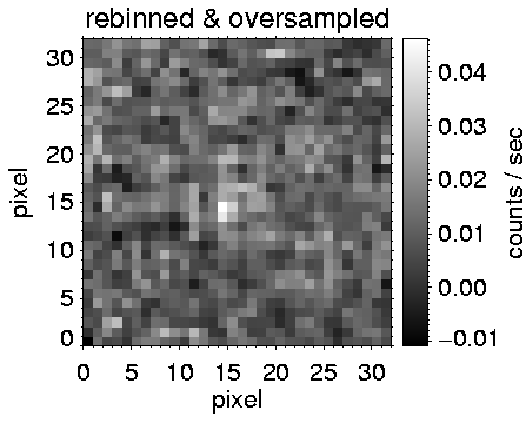}
     \end{minipage}} 
    \put(4.5,0.6){\begin{minipage}[t]{6.9cm}
        \includegraphics[width=\textwidth]{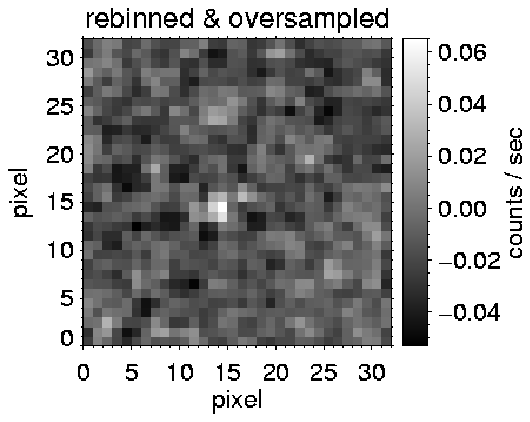}
     \end{minipage} }
   \put(0,0){ \framebox(14.5,18.5){}}
   \end{picture}
       \caption{\footnotesize{The reduced images for the L4.5 dwarf 2MASSW\,0337-1758. The top image represents the added image of all 4 exposures in filter F108N and the bottom one the residual image of all exposures after the SDI reduction (same as in Figures \protect \ref{Diff_imaging_result_1} and \ref{Diff_imaging_result_2}). In the middle panel, the left inlay displays the residual image at position 1 on the detector, while the right displays the residuals at position 2. The clearly remaining positive signal is visible in all three residual images and thus any contamination by a bad pixel can be ruled out. The brightness difference in the final image is $\Delta$\,F108N = 3.88\,mag. The orientation of all inlays is the same and the north direction is indicated by the arrow.}}
   \label{SDI_candidate_image}
  \end{figure}

The HST observations of the L4.5 brown dwarf 2MASSW\,033703-175807 (hereafter 2M0337-1758) revealed a clear residual signal in the final SDI reduced image, close to the center of the brown dwarf. Unlike for all other L dwarfs in this survey, this signal is also visible in the two individually reduced detector position images. Therefore, it cannot be caused by a single residual bad pixel. Further, with a S/N ratio of 7 at its peak in the final image, it is well above the noise level. Figure~\ref{SDI_candidate_image} displays the resulting images with the single detector position results in the middle and the final result, including all observation exposures, at the bottom.

\subsubsection{Evidence for binarity}
The F108N image of 2M0337-1758 and its final SDI reduced image were used to calculate the magnitude difference between the brown dwarf and the companion candidate, as well as their separation and position angle. This was achieved by applying aperture photometry with the IRAF \emph{phot} routine in the \emph{apphot} package. 
From the identified centroid positions of the photometric apertures a separation of 0.087$\arcsec$\,$\pm$\,0.015$\arcsec$ (corresponding to 2.52\,$\pm$\,0.44\,AU at a distance of 29\,pc) and a position angle of 195.6$\degr$ $\pm$ 5.5$\degr$ was calculated.

For the photometry, the preliminary assumption that the flux of the companion only contributes to the F108N filter and not the F113N filter was necessary, since the result of the applied SDI technique does not provide any information about the actual flux distribution of the companion candidate in each of the filters. This approximation is roughly valid for very late T dwarfs and even later spectral type objects (see Figure \ref{SDI_bandpasses}), but implies that the derived magnitude in F108N for the companion is only a lower limit. 
For further comparison with other observations and theoretical models, the instrumental count rates were converted into the Vega magnitude scale after aperture correction and using the most recent photometric keyword-value as provided by the STScI webpage\footnote{http://www.stsci.edu/hst/nicmos/performance/photometry/postncs\_keywords.html}.  Using this method, we derived a F108N magnitude of 16.54\,$\pm$\,0.02 mag for 2M\,0337-1758\,A, a brightness difference of 3.88\,$\pm$\,0.03 mag between the brown dwarf and its companion candidate and hence a F108N magnitude of 20.42\,$\pm$\,0.04 mag for the candidate. With a system distance of 29\,pc, this corresponds to absolute magnitudes of M$_{F108N}$\,=\,14.23\,$\pm$\,0.02 mag and 18.11\,$\pm$\,0.04 mag for A and B, respectively.

Together with the age estimation of the primary brown dwarf, the comparison of the derived values with the COND evolutionary models for brown dwarfs and EGPs (extrasolar giant planets) by \citet{Baraffe03} provides a first approximation of the mass and T$_{\mathrm{eff}}$ for the companion candidate. Since the F108N wavelength range lies much closer to the \emph{Y}-\,band ($\lambda_{\mathrm{C}}$\,=\,1.03\,$\mu$m) than to the \emph{J}\,-\,band, it was assumed that the objects are equally bright in F108N and \emph{Y}. For the following comparison with the isochrones, an extended version of the above mentioned evolutionary models including the \emph{Y}-\,band (I.\,Baraffe, private communication) was used. A comparison of the derived mass and effective temperature for the primary 2M0337-1758\,A with those one would derive with its 2MASS \emph{J} magnitude, proves the assumption of M$_{F108N}$ = M$_{Y}$ to be acceptable. The result is given in Table \ref{Mass_est_HST_comp}. For the companion candidate we derive a minimum mass of 10\,-\,15 $M_\mathrm{Jup}$ with a T$_{\mathrm{eff}}$ between $\sim$ 600\,K and $\sim$ 630\,K  for an age of 0.5 and 1 Gyr, respectively. To investigate how these results would change if the companion magnitudes in F108N and \emph{Y} are in fact not equally bright, we can assume two different scenarios as one can see in the right image of Figure~\ref{SDI_bandpasses}. In the two relevant bandpasses the flux of the T dwarfs peaks right in the F108N filter and these objects will therefore appear brighter in F108N than in the \emph{Y}-\,band. Assuming an offset of $\sim$\,0.5 mag for late T dwarfs, similar to the offset \citet{Biller06} derived between the \emph{H}-\,band and their SDI narrow band, the minimum mass of the companion would decrease down to 9\,-\,12 $M_\mathrm{Jup}$. In contrast, for a planetary mass object with 5 $M_\mathrm{Jup}$ at 1 Gyr the water absorption band extends into the F108N filter and this object would appear dimmer than in the \emph{Y}-\,band. Here, the offset might not be as large but to derive a valuable upper error limit, we use the same value. This results in an increase of the minimum mass to 11\,-\,17 $M_\mathrm{Jup}$ for ages of 0.5 and 1 Gyr, respectively. The results for the companion candidate are also summarized in Table \ref{Mass_est_HST_comp}, including the larger uncertainties due to the different assumptions.

As of fall 2009, there are only nine T dwarfs known, which are classified as T8 or later\footnote{http://www.DwarfArchives.org} and only five of them have a T$_{\mathrm{eff}}$ $\sim$ 600\,K (\citealt{Warren07}; \citealt{Delorme08}; \citealt{Burningham08}; \citealt{Burgasser08_2}; \citealt{Burningham09}). Hence, the companion adds up to a still very small number of just recently detected very cool brown dwarfs and the mass estimation places it right at the theoretical mass limit of $\sim$\,13 $M_\mathrm{Jup}$ for deuterium burning, which is mostly used as the boundary to distinguish between brown dwarfs and giant planets.
Nevertheless, the dominant uncertainty in translating the derived magnitude difference to brown dwarf and even planetary masses, is the uncertainty of the evolutionary models, which are not yet fully calibrated by observations, especially the mass-luminosity relation as a function of age (see further discussions below).

\subsubsection{A possible background object?}
To clarify if the detection might be caused by a background object, the position of 2M0337-1758 on previous HST/WFPC2 images was analyzed. These images were taken on 2000 December 25  in the broadband filter F814W and the medium band filter F1024M. No source could be detected in the WFPC2 frames above the noise level (corresponding to m$_{F814W}$\,=\,24\,mag) at the 2004 NICMOS position of the brown dwarf, but the comparison of the two positions allowed the first proper motion determination for 2M\,0337-1758 with $\mu_{\alpha}$\,cos\,$\delta$ = 0.247\,$\pm$\,0.02$\arcsec$/yr and $\mu_{\delta}$ = -\,0.024\,$\pm$\,0.028$\arcsec$/yr. This result will also be crucial to confirm the common proper motion nature of the binary in follow-up observations.

Analyzing the probability of a background source from a more physical side, we can already exclude a faint, stellar background source (like a M dwarf). These objects have, like L dwarfs, a relatively flat continuum between 1.08 -\,1.13\,$\mu$m and would be therefore not detected by our SDI technique. On the contrary, this method does not exclude a distant galaxy or a quasar. Based on the fact that we have a point-like remaining signal with an approximate magnitude of \emph{J}\,$\sim$\,19.5 mag, the candidate is unlikely a ``normal" galaxy and even too bright to be an active galactic nuclei (AGN). However, we cannot exclude a quasar or other peculiar background object (e.g.\,Wolf-Rayet galaxy). Therefore, the next step will be to establish common proper motion of the two components and to exclude any background object hypothesis.

\section{Discussion}
The direct detection of planetary mass objects still remains a challenging endeavor. In the past few years many systematic high-contrast direct imaging surveys for planetary mass companions around a large number of young, nearby stars have been conducted with HST (\citealt{Luhman05}; \citealt{Lowrance05}) and with AO at large ground\,-\,based telescopes (e.\,g.\,\citealt{Masciadri05}; \citealt{Metchev06}; \citealt{Lafreniere07}; \citealt{Biller07}; \citealt{Kasper07_2}, the latter two employing SDI). However, only very few planetary mass or very low mass brown dwarf companions have been discovered in these surveys. The first directly imaged planetary mass companion was detected around the young brown dwarf \object[2MASS J12073347-3932540]{2MASS\,1207-3932} (hereafter 2M1207) by \citet{Chauvin04, Chauvin05_01} with an estimated mass of 5\,-\,8 $M_\mathrm{Jup}$ and a separation of $\sim$ 55 AU. Shortly after, several more planetary mass object detections were reported to very young primaries with an age $\le$ 60 Myr: GQ Lupi\,b (\citealt{Neuh05}), AB Pic\,B (\citealt{Chauvin05_2}), DH Tau\,B (\citealt{Itoh05}; \citealt{Luhman06_2}), Oph\,1622-2405\,B (\citealt{RayJay06}; \citealt{Close07}), CHXR\,73\,B (\citealt{Luhman06_2}) and 1RXS\,J1609-2105\,B (\citealt{Lafreniere08}). With mass estimates between $\sim$\,7\,-\,20 $M_\mathrm{Jup}$ (thus at or just above the brown dwarf/planet boundary) and projected separations of 55\,-\,300 AU from their primary, their nature is still extensively debated and the exact number of planetary mass companions remains controversial. Such large separations suggest a formation process more similar to brown dwarfs (via fragmentation) rather than giant planets (via core accretion, out to a maximum distance of $\sim$ 10\,-\,15 AU; or via gravitational disk instability which demands a very high massive initial disk). Alternatively, planets could form close-in to the star and then migrate outward to a larger radii. \citet{MartinR07} show that such an outward migration could in principle be possible. 
Recently, \citet{Marois08} announced the first directly imaged multiple planetary system (HR\,8799\,b,c,d) with masses between $\sim$ 7\,-\,10 $M_\mathrm{Jup}$ at separations between 24 and 68 AU. At almost the same time, \citet{Kalas08} detected the so far lowest mass companion Fomalhaut\,b ($\sim$ 3 $M_\mathrm{Jup}$ ) and \citet{Lagrange09} claimed the discovery of the closest companion ever directly imaged to a star ($\beta$\,Pictoris\,b with $\sim$ 8 AU). While Fomalhaut\,b is again at a relatively large separation, HR\,8799\,b,c,d as well as $\beta$\,Pictoris\,b, are the first directly imaged companions around a star, whose formation could be explained by the common core accretion or disk instability scenarios.

But not only the formation process hinders the exact assignment of all these companions to the planetary regime. So far, the masses of these planet candidates are entirely based on evolutionary models. These models have been developed in significant detail over the past years (\citealt{Burrows97, Burrows05}; \citealt{Baraffe03, Baraffe08}) but remain purely theoretical, especially for planets and very young ages due to the lack of direct detections of well\,-\,characterized giant exoplanets. For example, the original mass estimate for GQ Lup\,b ranged from 1 to 42 $M_\mathrm{Jup}$ (\citealt{Neuh05}). This wide range in mass estimate is primarily caused by the use of different theoretical evolutionary models (with different initial conditions, see \citealt{Burrows97, Baraffe02, Wuchterl03}) to determine the mass as a function of luminosity, temperature and estimated age. The resulting huge discrepancy shows the difficulty of determining an absolute mass for such directly imaged low mass objects and especially at such young ages. 

\begin{figure}[ht]
  \centering
    \setlength{\unitlength}{1cm}
    \begin{minipage}[t]{8cm} 
       \includegraphics[angle=90, width=\textwidth]{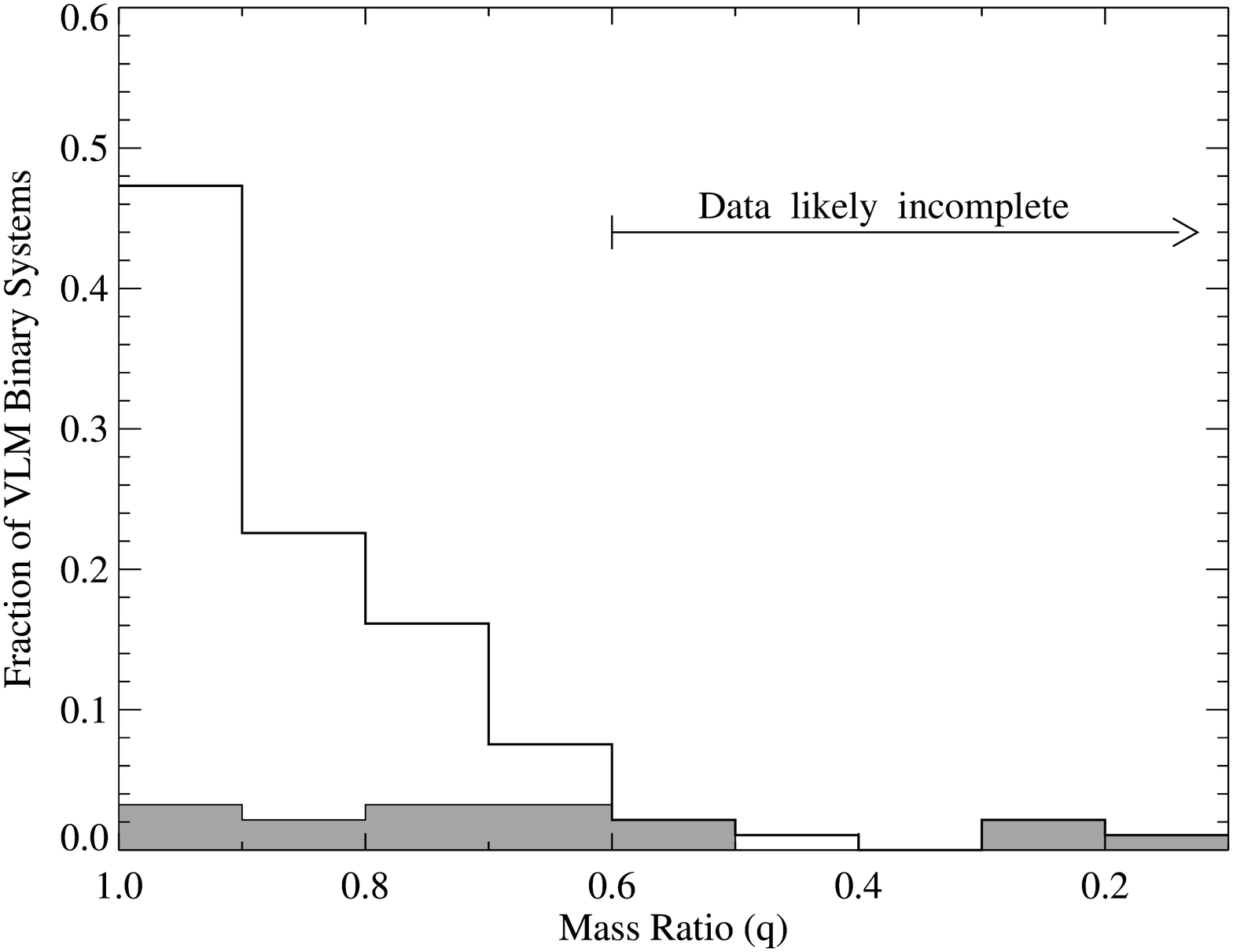}
      \end{minipage}\hfill
    \begin{minipage}[t]{7.8cm}
       \includegraphics[angle=90, width=\textwidth]{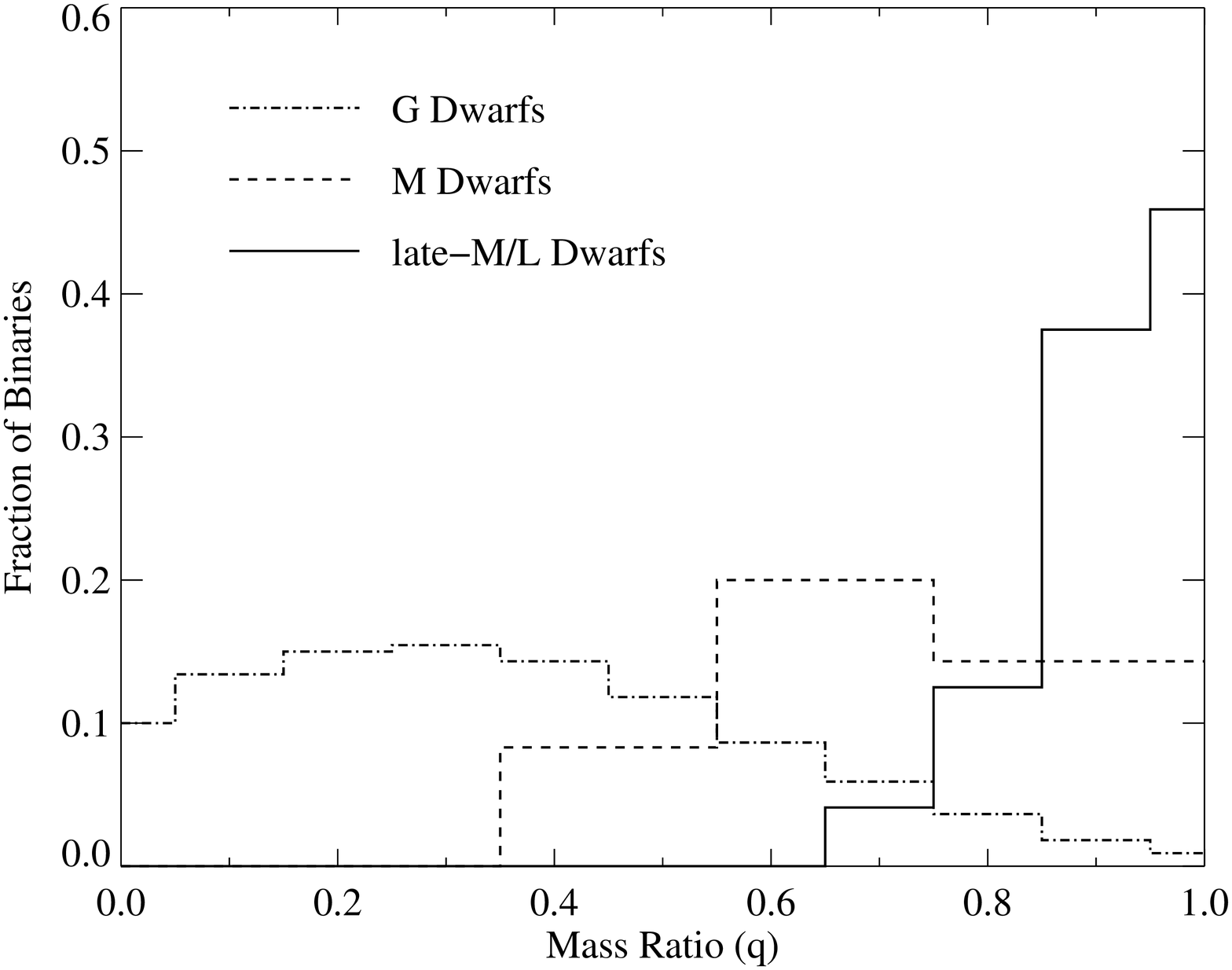}
     \end{minipage}
  \caption{ \footnotesize{The left image illustrates the mass ratio distribution computed for the currently known VLM binary systems, based on the online VLM Binaries Archive (www.vlmbinaries.org) as of July 2009. Spec\-tros\-copic binaries without known mass ratios were excluded. The distribution shows clear evidence of a peak near unity. The shaded bins represent the systems with age estimates $\le$\,10 Myr, which have on average larger separations compared to the older binary systems. The right image  displays the comparison of mass ratio distributions for G stars, early M stars and ultracool dwarfs, as adopted from \citet{Allen07}. The mass ratio appears to increas with decreasing mass of the primary.}}
   \label{mass_ratio_dist}
  \end{figure}

As mentioned in \S\,\ref{Motiv_SDI}, the first surveys for brown dwarf binaries already revealed that the resolved systems showed a tendency toward a high mass ratio distribution (M$_{2}$/M$_{1}$\,=\,\emph{q} $\ge$ 0.8) with a clear peak at \emph{q}\,$\sim$\,1. Further high-resolution surveys on ultracool dwarfs searching for binary systems strengthened this trend (see Figure\,\ref{mass_ratio_dist} on the left). Although an underestimation of the number of low mass ratio systems cannot be excluded due to the lack of sensitivity for \emph{q} $\le$ 0.5 in most of these surveys, the clear peak to equal mass binary systems is not an observational bias effect (for a detailed review see \citealt{Burgasser07_3} and for an extensive Bayesian study \citealt{Allen07}). The same surveys unfolded a semi-major axis distribution which peaks around 3\,-\,5\,AU with only very few separations beyond $\sim$ 15 AU (e.\,g.\,\citealt{Bouy03}; Table\,1 in \citealt{Burgasser07_3}; supported by numerical simulations of \citealt{Umbreit05} and \citealt{Allen07}).

In the following, the results of our HST survey will be compared with respect to the findings discussed above. The resolved and confirmed brown dwarf binary Kelu-1\,AB with its mass ratio \emph{q} = 0.82 and a separation of 5.59\,AU at the time of discovery, as well as 2MASSW\,0310+1648\,AB with its mass ratio \emph{q} $\sim$ 1 and a separation of 5.17 AU, are in full agreement with the previous findings and are an important contribution to the brown dwarf binary statistics. The non-detection of any companion in this survey at separations larger than 10 AU is also consistent with the separation statistics. The general observed binary fraction of very low mass stars (VLMS) and brown dwarfs, as found in direct imaging surveys is $\sim$\,15\,-\,20\%, and might be as high as $\sim$\,25\,-\,30\% (for extended discussions see e.\,g. \citealt{Burgasser07_3, Allen07}). With a total sample of 13 brown dwarfs, the two confirmed binary systems yield a binary fraction of $15^{+18}_{-10}$\,\% (or $23^{+19}_{-13}$\,\% if 2M0337-1758\,AB is confirmed as a third binary system\,\footnote{The uncertainty of the measured binary fraction was estimated by searching for the fraction of binaries at which the binomial distribution P$_B$(x,n,p) for x positive events in n trials falls off to 1/$e$ of its maximum value (cf. \citealt{Basri06}).}) for our survey. This is consistent with the results from previous VLMS/BD surveys. 

The discussion about the companion candidate to 2M0337-1758 can only be preliminary until second epoch observations confirm the companionship. Yet, it shows some very interesting properties. With a preliminary mass ratio of 0.18\,-\,0.22 (depending on its estimated age), this system would add to the very small number of resolved brown dwarf binaries with \emph{q} $\le$ 0.5. One of them, which has a resolved and confirmed  brown dwarf primary and a similar mass ratio is 2M1207\,AB (\emph{q} $\approx$ 0.3 if the most actual mass estimation of \citet{Mohanty07} is taken into account). But even if its secondary has a mass just below the deuterium burning limit, and therefore appears to be similar to the possible 2M0337-1758\,AB system, the \object[NAME 2M1207b]{2M1207\,AB} system has an unusual large separation and is much younger ($\sim$\,8 Myr). At these very young ages the surveys suggest a flatter mass ratio distribution than that for field or older cluster binaries (see shaded bins in Figure \ref{mass_ratio_dist} on the left), thus the low \emph{q} value might not be too unusual in the case of 2M1207\,AB. Another interesting system in this account is \object[SCR J1845-6357B]{SCR\,1845\,AB}. This M8.5\,-\,T6 field binary (where the secondary was discovered during a SDI survey as well) reveals also a  relatively low mass ratio with \emph{q} $\approx$ 0.41 (\citealt{Kasper07}). Additionally, the \object[SCR J1845-6357]{SCR\,1845\,AB} system is, with a separation of $\sim$\,4.5 AU, similar tightly bound as 2M0337-1758\,AB. Hence, both systems might be potential candidates for a dynamical mass determination on a reasonable time scale. This will further help to calibrate evolutionary models under a different aspect. The mass ratio distribution in binary systems exhibits a clear correlation with the primary mass (see Figure \ref{mass_ratio_dist} on the right). While G dwarf primaries show a relatively flat mass ratio distribution, the ratio for M dwarfs, VLMS and brown dwarfs peaks more towards unity (see also \citealt{Bergfors10} and references therein). Although a \emph{q} value of 0.18\,-\,0.22 is, independent of age and separation, still much larger than the mass ratios known from general star\,-\,giant planet systems (determined with RV, transit etc.\,methods), it does not completely exclude a brown dwarf\,-\,PMO system. Due to the lower initial mass of the primary the mass ratio of such a system would be a priori higher than for a main sequence star\,-\,giant planet system. Thus, such a \emph{q} value for our system probes an interesting ratio regime which thus far has not been explored for brown dwarfs. 

The minimum mass estimate of 10\,-\,15 $M_\mathrm{Jup}$ for the companion candidate of 2M0337-1758 does not a priori exclude a classification as a planet. Nevertheless, a comparison of four of the $\sim$\,600\,K  T dwarfs (all field dwarfs) in \citet{Leggett09} shows, that at least three of them (so far single objects) have similar mass estimates, just right below or above 13 $M_\mathrm{Jup}$. Therefore, 2M\,0337-1758 is more likely a brown dwarf -- brown dwarf binary system with a so far unusual very low mass secondary which possibly straddles the T/Y transition, rather than a system including an exoplanet.
Follow-up observations of the system are proposed to achieve a definite classification. Even more important, a larger number of further detections of objects in this mass and temperature regime are necessary to reliably calibrate the theoretical models. 

\section{Conclusions}
While previous direct imaging surveys focused especially on searches for planetary mass objects around young stars in nearby associations, we obtained high-resolution observations with HST/NICMOS of 12 L dwarfs within 30 pc of the sun, using the spectral differential imaging technique in the two narrow-band filters F108N and F113N. To this date, this is the only SDI survey for planetary mass companions around field brown dwarfs (not related to any young association or moving group) conducted from space or ground-based telescopes. Further, it is the only one which uses the spectroscopic feature of water absorption instead of the methane feature.

The survey resolved two brown dwarf binaries (Kelu-1\,AB and 2MASS\,0310+1648AB) for which common proper motion has already been confirmed in follow-up observations (\citealp{Stumpf09, Stumpf10_1}). The overall binary fraction of $15^{+18}_{-10}$\,\%, as well as separations of the binaries smaller than 6 AU are consistent with previous brown dwarf binary statistics. The mass ratios of \emph{q} $\ge$\,0.8 confirm the preference for equal mass systems as found in previous direct imaging surveys.

Furthermore, tentative evidence was found for a very low mass companion close to the planetary mass regime around 2MASS\,0337-1758. It has a very low T$_{\mathrm{eff}}$ $\approx$ 600\,-\,630 K and an unusual mass ratio (\emph{q} $\approx$ 0.2) compared to the vast majority of previously found brown dwarf binaries. Follow-up observations are proposed to confirm common proper motion of the system and to derive more physical properties like colors for spectral type classification. In addition, these new observations will provide second epoch astrometry for the determination of orbital parameters and hence an initial dynamical mass estimation. If gravitationally bound and depending on the age, the 2MASSW\,0337-1758 system will be one of the closest resolved brown dwarf binaries (0.087$\arcsec \pm$\,0.015$\arcsec$, corresponding to 2.52\,$\pm$\,0.44\,AU at a distance of 29\,pc) with the so far coldest companion ever directly imaged and the least massive companion to any L or T dwarf. 
Therefore, including its multiplicity status, it will then be an important testbed in the newly explored $\le$ 700 K temperature regime and might imply new constraints on the existing formation scenarios.


\acknowledgments

M.B.Stumpf and W. Brandner acknowledge support by the \emph{DLR Verbundforschung} project numbers 50 OR 0401 and 50 OR 0902.
We are grateful to Tricia Royle at STScI for her great and efficient support before and during observations. We would like to thank the anonymous referee for very helpful comments to improve the paper.
This research has benefitted from the M, L and T dwarf compendium housed at DwarfArchives.org and maintained by Chris Gelino, Davy Kirkpatrick and Adam Burgasser. This publication made also use of the Very-Low-Mass Binaries Archive housed at http://www.vlmbinaries.org and maintained by Nick Siegler, Chris Gelino and Adam Burgasser; and it made use of the SIMBAD database, operated at CDS, Strasbourg, France.

{\it Facilities:}  \facility{HST (NICMOS)}.

\clearpage


%
\clearpage

\clearpage
\centering
\begin{minipage}[t]{6.9cm}
        \includegraphics[width=\textwidth]{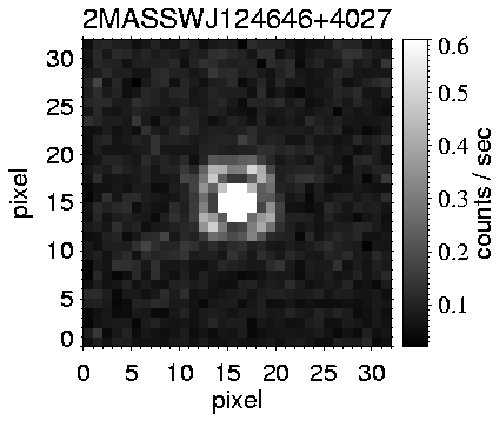}
    \end{minipage}\hfill
    \begin{minipage}[t]{6.9cm}
        \includegraphics[width=\textwidth]{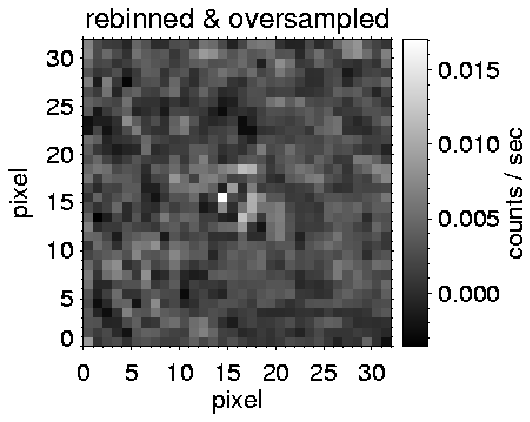}
     \end{minipage}\vspace{1.5ex}
   \begin{minipage}[t]{6.9cm}
        \includegraphics[width=\textwidth]{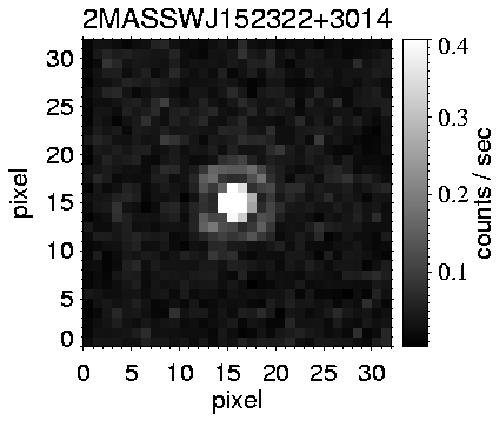}
    \end{minipage}\hfill
    \begin{minipage}[t]{7cm}
        \includegraphics[width=\textwidth]{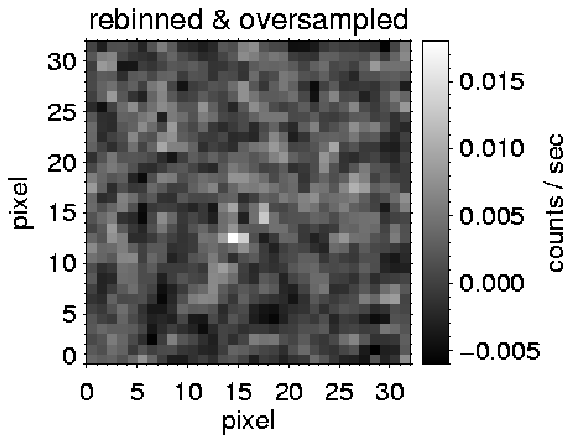}
     \end{minipage}
\centerline{Fig. 2. --- Continued.}
%

\clearpage

\clearpage

%
\clearpage

%
\clearpage

%

\clearpage

%

\clearpage

%

\clearpage

\begin{deluxetable}{l c c c c c c c c c}
\tabletypesize{\scriptsize}
\rotate
\tablecaption{List of selected Targets \label{Diff_imag_targets}}

\tablehead{
\colhead{Object name} & \colhead{RA} & \colhead{DEC} & \colhead{SpT} &
\colhead{Distance\,$^{\mathit{a}}$} & \colhead{\ion{Li}{1} EW} & \colhead{V\,$^{\mathit{b,d}}$} & \colhead{J\,$^{\mathit{c,d}}$} & \colhead{Observation} & \colhead{Ref.}\\
\colhead{} & \colhead{(J2000)} & \colhead{(J2000)} & \colhead{} & \colhead{[ pc ]} & \colhead{[ \r{A} ]} & \colhead{[ mag ]} & \colhead{[ mag ]} & \colhead{Date} & \colhead{}
}
\startdata
\object[2MASS J00452143+1634446]{2MASSW\,004521+1634} &  00$^\mathrm{h}$45$^\mathrm{m}$21.4$^\mathrm{s}$ & +16\degr34\arcmin44.7\arcsec & L3.5 & 10.4 & ? & 22.0 & 13.1 &   2004 Nov 17 &  3,4  \\
\object[2MASS J00511078-1544169]{2MASSW\,005110--1544} & 00$^\mathrm{h}$51$^\mathrm{m}$10.8$^\mathrm{s}$ & --\,15\degr44\arcmin16.9\arcsec & L3.5 & 30 & 10 & 24.1 & 15.2 & 2004 Dec 02 &  1 \\
\object[2MASS J01033203+1935361]{2MASSW\,010332+1935} &  01$^\mathrm{h}$03$^\mathrm{m}$32.0$^\mathrm{s}$ &  +19\degr35\arcmin36.2\arcsec & L6 & 28.6 & 12 & 24.5 & 16.1 & 2004 Oct 08 &  1 \\
\object[2MASS J03105986+1648155]{2MASSW\,031059+1648} & 03$^\mathrm{h}$10$^\mathrm{m}$59.9$^\mathrm{s}$ & +16\degr48\arcmin15.6\arcsec & L9 & 20 & 5 & 24.8 & 16.4 & 2004 Sep 24 &  1   \\
\object[2MASS J03370359-1758079]{2MASSW\,033703--1758} & 03$^\mathrm{h}$37$^\mathrm{m}$03.6$^\mathrm{s}$ & --\,17\degr58\arcmin07.9\arcsec & L4.5 & 29 & 8 & 23.6 & 15.6 & 2004 Sep 24 &  1 \\
\object[LSR J0602+3910]{LSR\,0602+3910} & 06$^\mathrm{h}$02$^\mathrm{m}$30.5$^\mathrm{s}$ & +39\degr10\arcmin59.2\arcsec & L1 & 10.6 & 7 & 20.8 & 12.3 & 2005 Apr 15 &  4 \\
\object[2MASS J06523073+4710348]{2MASSI 065230+4710} & 06$^\mathrm{h}$52$^\mathrm{m}$30.7$^\mathrm{s}$ & +47\degr10\arcmin34.8\arcsec & L4.5 & 11.1 & 14 & 21.4 & 13.5 &  2005 Mar 24& 5 \\
\object[2MASS J08251968+2115521]{2MASSI 082519+2115} & 08$^\mathrm{h}$25$^\mathrm{m}$19.6$^\mathrm{s}$ & +21\degr15\arcmin52.0\arcsec & L7.5 & 10.7\,$^{*}$ & 10 & 22.5 & 15.1 & 2005 Feb 12 & 1,2  \\
\object[2MASS J08354256-0819237]{2MASSI 083542--0819} & 08$^\mathrm{h}$35$^\mathrm{m}$42.6$^\mathrm{s}$ & --\,08\degr19\arcmin23.7\arcsec & L5 & 8.3 & ? & 21.0 & 13.2 & 2005 Jun 14 &  5 \\
\object[2MASS J12464678+4027150]{2MASSW\,124646+4027} & 12$^\mathrm{h}$46$^\mathrm{m}$46.8$^\mathrm{s}$ & +40\degr27\arcmin15.1\arcsec & L4 & 25.1 & 11 & 22.3 & 15.0 & 2004 Oct 22 &  1  \\
\object[2MASS J13054019-2541059]{Kelu-1} & 13$^\mathrm{h}$05$^\mathrm{m}$40.2$^\mathrm{s}$ & --\,25\degr41\arcmin06.0\arcsec & L2 & 18.7\,$^{*}$ & 1.7 & 22.1 & 13.4 & 2005 Jul 31&  2,6 \\
\object[2MASS J15232263+3014562]{2MASSW\,152322+3014} & 15$^\mathrm{h}$23$^\mathrm{m}$22.6$^\mathrm{s}$ & +30\degr14\arcmin56.2\arcsec  & L8 & 18.6\,$^{*}$ & 9 & 24.4 & 16.3 & 2004 Sep 30 &  1  \\
\noalign{\medskip}
\object[2MASS J05591914-1404488]{2MASSW\,055919--1404} & 05$^\mathrm{h}$59$^\mathrm{m}$19.1$^\mathrm{s}$ &  --\,14\degr04\arcmin48.9\arcsec & T4.5 & 10.3 & -- & 25.0 & 13.8 & 2004 Sep 07 & 2,7 \\  
\enddata
\tablenotetext{?}{\, denotes lacking observations}
\tablenotetext{-}{\, no observation since atomic lithium forms into LiCl for T$_{\mathrm{eff}}$\,$\le$\,1400\,K \citep{Burrows99} and objects that cold are brown dwarfs by default}
\tablenotetext{a}{\,Spectrophotometric distance estimates from the given references unless otherwise noted. Distance error is $\sim$ 10\%} 
\tablenotetext{*}{\,Distance from trigonometric parallax in the given references}
\tablenotetext{b}{\,from the CDS Simbad service}
\tablenotetext{c}{\,from 2MASS All-Sky Point Source Cataloge}
\tablenotetext{d}{\,Uncertainties of apparent magnitudes are $\sim$ 0.1\,mag}
\tablerefs{
(1)\,\citet{Kirk00}, 
(2)\,\citet{Dahn},
(3)\,\citet{Wilson03},
(4)\,\citet{Salim03},
(5)\,\citet{Cruz03},
(6)\,\citet{Kirk99},
(7)\,\citet{Burgasser03_2}}

\end{deluxetable}

\clearpage

\begin{deluxetable}{l c c c c c  }
\tablecaption{Mass estimations for the companion candidate of 2MASSW\,0337-1758 \label{Mass_est_HST_comp}}
\tablehead{
\colhead{}&\colhead{}& \multicolumn{2}{c}{0.5 Gyr}\hspace{2ex} & \multicolumn{2}{c}{1 Gyr}\hspace{2ex}\\ 
\colhead{\raisebox{2.4ex}[2.4ex]{Object}\hspace{0.5ex}} &\colhead{\raisebox{2.4ex}[2.4ex]{M$_{Y}$[\,mag\,]}\hspace{2ex}} & \colhead{$M_\mathrm{Jup}$} & \colhead{T$_{\mathrm{eff}}$ [\,K\,]\hspace{2ex}} & \colhead{$M_\mathrm{Jup}$} & \colhead{T$_{\mathrm{eff}}$ [\,K\,]} 
}
\startdata
Brown Dwarf\,$^{\mathit{a}}$& 14.23\,$\pm$\,0.02\hspace{2ex} & 54\,$\pm$\,2 & 1870\,$\pm$\,60 \hspace{2ex}& 68\,$\pm$\,1 & 1945\,$\pm$\,35 \\[1.5ex]
companion candidate\,$^{\mathit{b}}$ & 18.11\,$\pm$\,0.50\hspace{2ex} & $\ge$ 10 & 600\,$\pm$\,70 \hspace{2ex}& 15\,$\pm$\,2 & 630\,$\pm$\,70
\enddata
\tablecomments{Masses and effective temperatures derived from:}
\tablenotetext{a}{\,DUSTY00 evolutionary models by \citet{Chabrier00, Baraffe02}}
\tablenotetext{b}{\,COND03 evolutionary models by \citet{Baraffe03}}
\end{deluxetable}

\end{document}